\documentclass[bibyear]{aa}
\usepackage[varg]{txfonts}
\usepackage{graphicx}
\usepackage{longtable}
\usepackage{lscape}
\newcommand{\e}{et al.\ }
\newcommand{\ha}{H$\alpha$}
\newcommand{\teff}{T$_{eff}$}
\newcommand{\logg}{$\log g$}
\newcommand{\hii}{H$_{\rm II}$}
\begin{document}

\title{Gaia-ESO Survey: global properties of clusters Trumpler 14 and 16
in the Carina Nebula.
\thanks{
Based on observations collected with the FLAMES spectrograph at VLT/UT2
telescope (Paranal Observatory, ESO, Chile), for the Gaia-ESO Large
Public Survey (program 188.B-3002).
}
}

\date{Received date / Accepted date}

\author{F. Damiani\inst{1},
A. Klutsch\inst{2},
R.~D. Jeffries\inst{3},
S. Randich\inst{4},
L. Prisinzano\inst{1},
J. Ma\'iz Apell\'aniz\inst{5},
G. Micela\inst{1},
V. Kalari\inst{6,7},
A. Frasca\inst{2},
T. Zwitter\inst{8},
R. Bonito\inst{1,9},
G. Gilmore\inst{10},
E. Flaccomio\inst{1},
P. Francois\inst{11},
S. Koposov\inst{10,12},
A.~C. Lanzafame\inst{13},
G.~G. Sacco\inst{4},
A. Bayo\inst{14},
G. Carraro\inst{15},
A.~R. Casey\inst{10},
E.~J. Alfaro\inst{16},
M.~T. Costado\inst{16},
P. Donati\inst{17},
E. Franciosini\inst{4},
A. Hourihane\inst{10},
P. Jofr\'e\inst{10,18},
C. Lardo\inst{19},
J. Lewis\inst{10},
L. Magrini\inst{4},
L. Monaco\inst{20},
L. Morbidelli\inst{4},
C.~C. Worley\inst{10},
J. Vink\inst{6},
\and
S. Zaggia\inst{21}
}
\institute{INAF - Osservatorio Astronomico di Palermo G.S.Vaiana,
Piazza del Parlamento 1, I-90134 Palermo, Italy \\
\email{damiani@astropa.inaf.it}
\and
INAF - Osservatorio Astrofisico di Catania, via S. Sofia 78, 95123,
Catania, Italy
\and
Astrophysics Group, Keele University, Keele, Staffordshire ST5 5BG,
United Kingdom
\and
INAF - Osservatorio Astrofisico di Arcetri, Largo E. Fermi
5, 50125, Firenze, Italy
\and
Centro de Astrobiolog\'ia (CSIC-INTA), ESAC campus, Camino bajo del
castillo s/n, 28 692 Villanueva de la Ca\~nada, Madrid, Spain
\and
Armagh Observatory, College Hill, Armagh BT61 9DG, UK
\and
School of
Mathematics \& Physics, Queen's University Belfast, Belfast BT61 7NN, UK
\and
Faculty of Mathematics and Physics, University of Ljubljana, Jadranska
19, 1000, Ljubljana, Slovenia
\and
Dipartimento di Fisica e Chimica, Universit\`a di Palermo,
Piazza del Parlamento 1, 90134, Palermo, Italy
\and
Institute of Astronomy, University of Cambridge, Madingley Road,
Cambridge CB3 0HA, UK
\and
GEPI, Observatoire de Paris, CNRS, Universit\'e Paris Diderot, 5 Place
Jules Janssen, 92190 Meudon, France
\and
Moscow MV Lomonosov State University, Sternberg Astronomical Institute,
Moscow 119992, Russia
\and
Dipartimento di Fisica e Astronomia, Sezione Astrofisica, Universit\`{a}
di Catania, via S. Sofia 78, 95123, Catania, Italy
\and
Instituto de F\'isica y Astronomi\'ia, Universidad de Valparai\'iso, Chile
\and
European Southern Observatory, Alonso de Cordova 3107 Vitacura, Santiago
de Chile, Chile
\and
Instituto de Astrof\'isica de Andaluc\'ia-CSIC, Apdo. 3004, 18080,
Granada, Spain
\and
INAF - Osservatorio Astronomico di Bologna, via Ranzani 1, 40127,
Bologna, Italy
\and
N\'ucleo de Astronom\'ia, Facultad de Ingenier\'ia,
Universidad Diego Portales,  Av. Ejercito 441, Santiago, Chile
\and
Astrophysics Research Institute, Liverpool John Moores University, 146
Brownlow Hill, Liverpool L3 5RF, United Kingdom
\and
Departamento de Ciencias Fisicas, Universidad Andres Bello, Republica
220, Santiago, Chile
\and
INAF - Osservatorio Astronomico di Padova, Vicolo dell'Osservatorio 5,
I-35122, Padova, Italy
}

\abstract{}{
We present the first extensive spectroscopic study of the global
population in star clusters Trumpler~16, Trumpler~14 and Collinder~232
in the Carina Nebula, using data from the Gaia-ESO Survey, down to
solar-mass stars.}
{In addition to the standard homogeneous Survey data reduction,
a special processing was applied here because of
the bright nebulosity surrounding Carina stars.}
{We find about four hundred good candidate members
ranging from OB types down to slightly sub-solar masses.
About one-hundred heavily-reddened early-type Carina members found here were
previously unrecognized or poorly classified,
including two candidate O stars and several candidate Herbig Ae/Be stars.
Their large brightness makes them useful tracers of the
obscured Carina population. The
spectroscopically-derived temperatures for nearly 300 low-mass members
allows the inference of
individual extinction values, and the study of the relative placement of
stars along the line of sight.}
{We find a complex spatial structure, with definite
clustering of low-mass members around the most massive stars,
and spatially-variable extinction.
By combining the new data with existing X-ray data
we obtain a more complete picture of the
three-dimensional spatial structure of the Carina clusters, and of their
connection to bright and dark nebulosity, and UV sources.
The identification of tens of background giants enables us also to determine
the total optical depth of the Carina nebula along many sightlines.
We are also able to put constraints on the star-formation history of
the region, with Trumpler~14 stars found to be systematically younger
than stars in other sub-clusters.
We find a large percentage of
fast-rotating stars among Carina solar-mass members, which provide new
constraints on the rotational evolution of pre-main-sequence stars in
this mass range.
}


\titlerunning{Trumpler 14 and 16 in Carina Nebula}
\authorrunning{Damiani et al.}
\maketitle

\section{Introduction}
\label{intro}

The Carina Nebula is one of the most massive \hii\ regions known in the
Galaxy. It contains a large population of massive OB stars (the Car OB1
association), several
Wolf-Rayet stars, and the well known LBV star $\eta$~Carinae. Most of
the stellar content of the Carina Nebula is found concentrated in a few
clusters, notably Trumpler~16 (Tr~16, hosting $\eta$~Car itself) and
Trumpler~14 (Tr~14), about 10$^{\prime}$ NNW of $\eta$~Car. Less conspicuous
clusters associated with the Nebula include Trumpler~15, Collinder~228
and Collinder~232. The distance to $\eta$~Car has been precisely
determined to be 2.35$\pm 0.05$~kpc (Smith 2006).
Car OB1 contains some of the most massive O stars known, including rare
examples of O3, and even O2 stars. The properties of the region were
reviewed by Smith and Brooks (2007, 2008). More recently, the whole
Carina star-formation region (SFR) was thoroughly investigated by means
of a mosaic of Chandra X-ray observations (CCCP: Chandra Carina Complex
Project; Townsley \e 2011, and all papers
in the series), after earlier X-ray studies of the central clusters
Trumpler~16 and~14 with both Chandra (Albacete-Colombo \e 2008) and
XMM-Newton (Antokhin \e 2008). The X-ray data have been crucial to
demonstrate the existence of a population (both clustered and diffuse)
of $\geq 14000$ stars, undoubtedly associated with the SFR, being most
likely low-mass young stars formed in the Nebula several millions years
ago. Detailed studies of the stellar population in Carina have been
until now exclusively directed towards characterizing its rich
massive-star members, while are still largely missing for its lower-mass
population. For example, DeGioia-Eastwood \e (2001) reported optical photometry
for only $\sim 850$ stars in Trumpler~16 and~14.
Only fairly recently deep optical photometry on more than 4500 stars in
the same region was published by Hur \e (2012), allowing optical counterparts
of faint X-ray sources to be studied.
Spectroscopic studies of the low-mass PMS stars in these clusters are
almost entirely lacking; Vaidya \e (2015) present low-resolution spectra
of 11 PMS stars.

The study of low-mass PMS stars at the distance of Carina, and sometimes
embedded within obscuring dust and/or bright nebular emission, is
time-consuming and technically challenging.
At the same time, it is important to test whether the early evolution of
stars under such ``extreme'' ambient conditions, dominated by the
presence of hundreds OB stars, do differ from that in ``quieter'' SFRs
(e.g.\ Taurus-Auriga, or also Orion).
Recent results from X-ray and IR surveys suggest that stars formed in
rich clusters (e.g., Carina, Cygnus~X, NGC3603, Westerlund~1 and~2) may
be an important, if not dominant, component of all stars in the Galaxy,
thus more representative of the ``average Milky-Way star'' than stars formed in
less rich SFRs like e.g., Tau-Aur, Chamaeleon, or IC348
(see e.g., Lada and Lada 2003). Therefore, the study of Carina stars
across the whole mass spectrum is likely to be relevant for a better
understanding of the general stellar population in the Galaxy.

It is not clear whether the various clusters in Carina
are coeval, and if not, if this can be attributed to triggered or
sequential star
formation processes. Evidences for triggered formation in Carina have been
discussed by Smith \e (2010), but focusing on a different part of the
nebula (the 'southern pillars') than that studied here, with some
overlap only in the Tr~16 SE obscured region.
In the central part of Carina studied here, evidence for
recent or ongoing star formation is less frequent than in the southern parts
(Povich \e 2011b).
Among the central Carina clusters, Tr~14 was suggested to be 1-2 Myr
younger than Tr~16, because of its more compact structure and other
characteristics (Walborn 1995, Smith 2006). It is however unclear if
Tr~16 can still be
considered as a single cluster with a rather sparse population, or rather
as several physical clusters, as suggested by the X-ray results of Feigelson \e
(2011).

In order to complement the studies of massive stars in Carina, and
obtain a more complete understanding of the star formation processes which
have taken place in its recent past, we present here the first
spectroscopic study of a sizable population of hundreds lower-mass stars
(down to approximately one solar mass), using data from the
Gaia-ESO Survey (Gilmore \e 2012, Randich \e 2013).
The same observational dataset was used in a previous work (Damiani \e
2016, Paper~I) to study the dynamics of the ionized gas in the Carina \hii\
region from optical nebular emission lines.

This paper is structured as follows: in Section~\ref{obs} we describe
the composition of the observed sample;
Section~\ref{membership} discusses cluster membership for the observed stars;
Section~\ref{ob} discusses massive stars which happen to fall within our sample;
Section~\ref{spatial} discusses the spatial clustering of stars;
Section~\ref{reddening} is devoted to a discussion of reddening;
Section~\ref{cmd} presents Color-magnitude diagrams;
Section~\ref{xray} compares results from the present data with those
from X-ray data;
Section~\ref{structure} discusses the structure of the whole region;
Section~\ref{ages} discusses stellar ages, and Section~\ref{rotation}
the rotational properties of Carina stars.
{ Eventually, we summarize our results in Section~\ref{concl}.}

\section{Observations and data analysis}
\label{obs}

\begin{figure}
\resizebox{\hsize}{!}{
\includegraphics[bb=5 10 485 475]{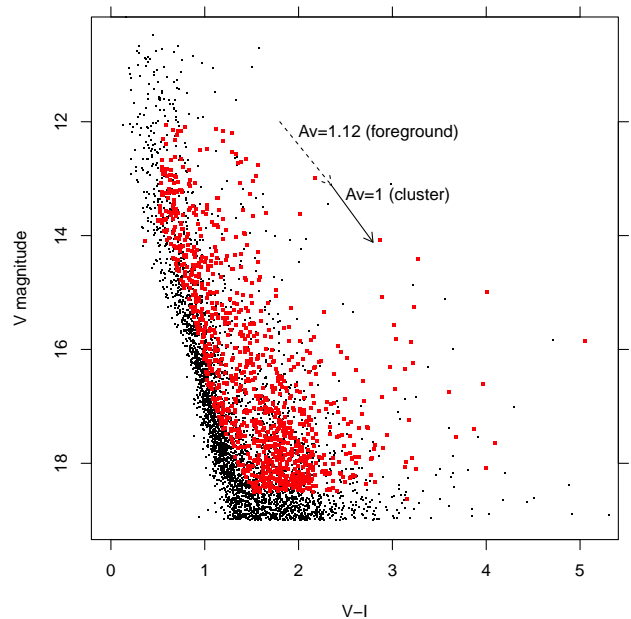}}
\caption{$(V,V-I)$ diagram of all stars in the Hur \e (2012) catalog.
Big red dots indicate the spectroscopically observed stars studied here.
The dashed arrow indicates the foreground reddening vector, while
the solid arrow indicate a representative intra-cluster reddening vector
corresponding to $A_V=1$.
\label{targ-sel-v-vi}}
\end{figure}

\begin{figure}
\resizebox{\hsize}{!}{
\includegraphics[bb=5 10 485 475]{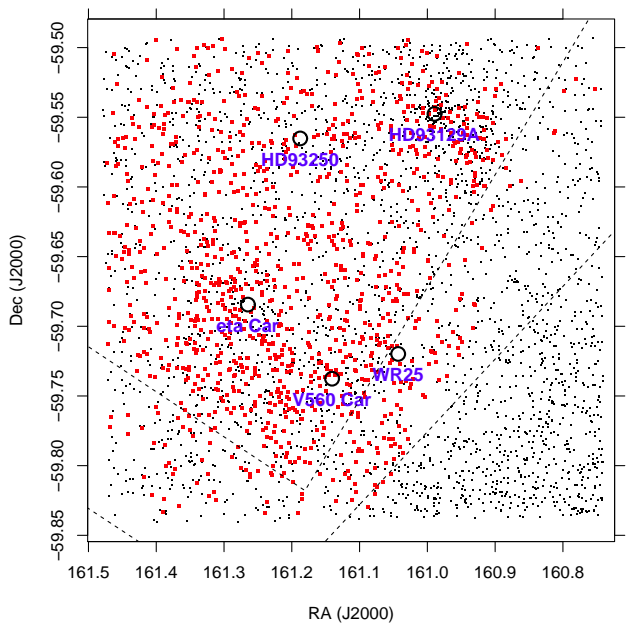}}
\caption{Spatial distribution of Hur \e (2012) stars brighter than
$V=18.5$, and of observed stars (red).
The most massive stars in the region are indicated with big circles and names.
The dashed lines indicate the approximate positions of the dark obscuring
dust lanes bounding Tr~14 and~16.
\label{targ-sel-ra-dec}}
\end{figure}

\subsection{The observed sample}
\label{sample}

In this work we study a set of spectra of 1085 distinct stars in the
Carina Nebula, obtained with the FLAMES/Giraffe multi-fibre spectrometer at the
ESO VLT/UT2 telescope on April 6-9, 2014,
and released as part of the iDR4 Gaia-ESO data release.
The total number of individual spectra was 1465,
but all spectra relative to the same star were coadded to improve
signal-to-noise ratio (SNR).
Simultaneously, spectra with the FLAMES/UVES high-resolution
spectrograph were obtained, which will be presented separately (Spina \e in
prep.).
All spectra considered here are taken using Giraffe setup
HR15n ($R \sim 17000$, wavelength range 6444-6818\AA), as in all
Gaia-ESO observations of cool stars in open clusters; the (known) OB stars in
Carina are instead being observed using different Giraffe setups, to
obtain a more comprehensive set of diagnostics, better suited to hot
stars (the massive star spectra are not part of the current data release
and will be studied in a later work).
As a consequence, we will be able to give only a rough
classification for the early-type stars which happen to fall within the
sample studied here. The approach for deriving stellar parameters for
later spectral types was instead described in Damiani \e (2014) and
Lanzafame \e (2015).

The basis for our FLAMES target selection was the
optical photometry published by Hur \e (2012). This study covered a
field, approximately $25^{\prime} \times 25^{\prime}$ in size, comprising
only the central portions of the entire Nebula, and thus
limited to the large clusters Tr~16 and~14, and
the less rich Collinder~232. Moreover, we have not observed with the
HR15n setup stars brighter than $V=12$ (massive stars, observed
with different Giraffe setups), nor fainter than $V=18.5$, in order to obtain
an acceptable minimum SNR. Since low-mass stars in the Carina SFR have not
yet arrived on the Zero-Age Main Sequence (ZAMS),
we also excluded stars in that region of the
color-magnitude diagram (CMD); also a spatial region near the edge of
the photometric field-of-view, apparently dominated by field stars, was
not observed; finally, random sampling of the remaining stars was made
to avoid an exceedingly long target list. This procedure follows the
general strategy for target selection in the Gaia-ESO Survey described
in Bragaglia \e (in preparation).
The CMD of both the Hur \e (2012) input catalog and our
spectroscopically observed stars is shown in
Fig.~\ref{targ-sel-v-vi}. Fig.~\ref{targ-sel-ra-dec} shows instead the
spatial distribution of input and observed stars.
The median SNR of our 1085 spectra is 36.2.

\subsection{Data analysis}
\label{reduction}

The Gaia-ESO Survey data analysis process is distributed among several
Working Groups, whose task is to apply homogeneous procedures to all
datasets to ensure a high degree of internal coherence. As explained in
Paper~I, however, the Carina Nebula presents a number of unanticipated
features, which are not dealt with accurately using standard
procedures/pipelines, like the spatially nonuniform sky continuum level
due to reflection nebulosity and the wide \ha\ wings due to both
high-velocity ionized gas and reflection in the Nebula. This required to
take a step back and redo part of the analysis, especially regarding a
more appropriate subtraction of the sky spectrum, as described in the
following. All derived parameters used here are released in the internal
{\em GESiDR4} data release, database table {\em AstroAnalysis}. This
reports the detailed results from each analysis ``node'', and their
merged values produced by the relevant Working Group (WG12 in our case);
because of the mentioned difficulties, results obtained from individual nodes
sometimes using non-standard, ad-hoc procedures
were preferred to WG12 results in the case of stellar parameters (from
node ``OAPA'') and of lithium equivalent widths (EW; node ``OACT''),
while radial
and rotational velocities were taken as those produced by WG12 (or node
``OAPA'' when these latter were missing).

The details of the special background-subtraction procedure employed by the
OAPA node before evaluating stellar parameters are given in
Appendix~\ref{append}. In short, after having separately corrected all
atmospheric sky features, nebular lines are corrected using each of the nearest
five sky spectra: the range of corrections obtained is an estimate of
the uncertainty involved in the procedure, while the median of the five
corrected spectra is taken as the best estimate stellar spectrum, for
stellar parameter derivation. Lithium EWs were computed by different
nodes, with results in good mutual agreement, using slightly
different methods as explained in Lanzafame \e (2015); the EW set from
the OACT node was chosen because of its largest sample coverage.
The WG12 radial and rotational velocities include contributions from
both the OACT node, using methods detailed in Frasca \e (2015), and OAPA
node, using cross-correlation as briefly described in Damiani \e (2014);
a very good agreement is found for stars in common.

With the collection of five-fold nebular-subtracted stellar
spectra obtained as explained in Appendix~\ref{append},
we proceeded with our estimates of stellar
parameters (done five times per star), using the method explained in
Damiani \e (2014).
{ This latter was purposely developed to deal with the spectral range and
resolution of Giraffe HR15n data, and defines a set of spectral indices
and their calibration to derive \teff, \logg, and [Fe/H] for stars later
than $\sim$A2.}
The range spanned by each parameter
in its own set of five determinations corresponds to the systematic
error introduced by sky correction, often larger than the
statistical error.
Depending on the
intensity ratio between the nebular \ha\ wings and and the stellar \ha\
wings, these latter may in some cases be exceedingly affected by
nebular emission, and cannot be used as a temperature diagnostic, e.g. in
A to mid-G stars\footnote{The most important features of stellar spectra
in the HR15N wavelength range are described in Damiani \e (2014).}.
For mid-A to lower-mass stars, fortunately, other
indicators may be used to estimate temperatures (although the increased
dependence on metallicity must be treated with caution).  The problem
is worst for early-A stars, where almost the only spectral line in HR15n
range is \ha. Much weaker lines are found in the blue extreme of the HR15n
range, where grating efficiency is however very low, and are therefore
difficult to use in low-SNR spectra.  In practice, many faint early-A
stars are recognized in the Tr~14/16 dataset as those having a nearly flat,
featureless continuum spectrum, with indefinite properties in the \ha\
region because of the predominant nebular emission. B-type stars are
instead easily recognized because of their He~I 6678\AA\ line: the He~I
nebular line at the same wavelength has usually no wings (Paper~I), and is much
narrower than the stellar line, often broadened by fast
rotation.

The wide nebular \ha\ wings are also of great nuisance when trying to
selecting stars with intrinsic wide \ha\ emission because of
accretion from a circumstellar disk (e.g.\ Classical T Tauri stars - CTTS - or
Herbig Ae/Be stars - HAeBe).
Here again, the five-fold nebular-subtracted star spectra are of invaluable
help in separating cases where the apparent wide emission in the net spectrum
arises from poorly subtracted nebular \ha\ wings (as it will not be
present in all five net spectra)
from truly wide \ha\ lines of CTTS/HAeBe stars.

\subsection{Auxiliary data sets}
\label{aux}

The stellar population of the Carina nebula was the object of several
recent studies. We have therefore cross-matched our spectroscopic
targets with objects in several existing catalogues.
A match with the CCCP X-ray source catalog (Broos \e 2011a) yielded 352
matches among our 1085 sample stars. The match was made assuming a
constant $1\sigma$ error on optical positions of 0.2~arcsec,
individual catalogued X-ray position errors,
and a 4-$\sigma$
maximum distance. The number of spurious matches was estimated as 13, by
artificially shifting one of the two position lists by $\pm 1$~arcmin.
Then we considered the VPHAS$+$ DR2 Point Source Catalogue (Drew \e
2014) with photometry in the bands $ugri$ and H$\alpha$.
Using a maximum matching distance of 5~$\sigma$, and constant position
errors of 0.2~arcsec for both optical and VPHAS$+$ catalogues, we obtain
1074 matches (of which $\sim 140$ estimated as spurious).
However, the
number of stars with a {\it clean=yes} photometric flag in all of $r$, $i$, and
H$\alpha$ bands, matching our target list, is of only 171.
Last, we matched similarly our targets to the Young Stellar Object (YSO)
catalogue of Zeidler \e
(2016), obtained from both VISTA and Spitzer near/mid-IR data.
The number of matches is 64 ($\sim 1$ spurious), of
which however only 7 have catalogued magnitudes in all four Spitzer IRAC
bands ($3.6-8.0\mu$). All of these latter were already members by
spectroscopic criteria, see below.

\section{Cluster membership}
\label{membership}

Since the chosen sample-selection strategy for Gaia-ESO Giraffe observations is
inclusive of most possible members, membership of observed stars to each
cluster must be determined post-facto.
The emphasis here being on young, low-mass stars (FGK stars), the most
useful membership indicators are the lithium line 
EW and strong X-ray emission, in addition to radial velocity
(RV) as usual. Wide \ha\ emission wings, indicative of circumstellar
accretion in pre-main-sequence (PMS) stars,
is a strong evidence of extreme youth and probable membership to
a SFR, even in the absence of a RV measurement (e.g., in cases of
strong-emission stars without observable photospheric absorption lines).
As described above, a star must show wide \ha\ emission consistently
across the multiple (five) sky-subtraction options, in order to be
considered a reliable CTTS/HAeBe. This conservative approach undoubtedly
misses some real emission-line members, but is a minor issue for
building a reliable member list, since the other membership criteria may
compensate for this.
Narrow, chromospheric \ha\ emission cannot be used here as a youth
indicator, being swamped by the nebular emission lines whatever the
sky-subtraction option chosen.
In order to minimize the number of spurious members, a star is generally
accepted as member only when it satisfies at least two criteria among:
{ $|RV-RV_0| < \Delta RV$ (with $RV_0$ and $\Delta RV$ to be determined),}
lithium EW$>150$~m\AA, X-ray detection, and CTTS/HAeBe status.
However, since each of these criteria may include significant numbers of
contaminants (e.g., G stars up to an age of about 1~Gyr would fall above
our lithium EW threshold),
some additional screening was applied, as described below.

\begin{figure}
\resizebox{\hsize}{!}{
\includegraphics[bb=5 10 485 475]{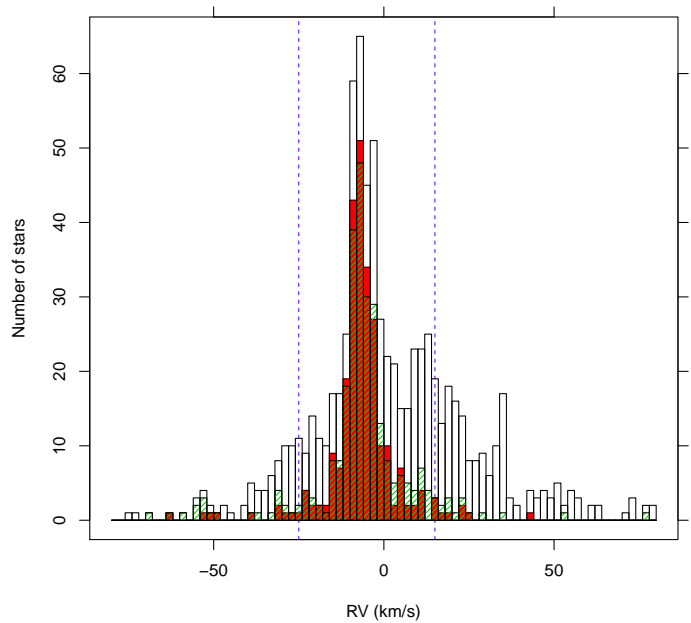}}
\caption{Histogram of { (heliocentric)} RV for all stars with
$T_{eff}<8800$~K and SNR$>15$
(white-colored bars).  The red filled histogram represents stars with
lithium EW$>150$~m\AA; the dashed green histogram are stars detected in
X-rays. The vertical dashed blue lines at $RV=(-25,+15)$ indicate our
fiducial limits for RV membership.
\label{rv-hist}}
\end{figure}

{ To help determining $RV_0$ and $\Delta RV$, Figure~\ref{rv-hist} shows
a histogram of RV for stars with $T_{eff}<8800$~K (and
SNR$>15$).
Colored histograms refer to lithium- and X-ray-selected candidates: they
select essentially the same population of low-mass Carina members.
From this Figure, we find that good values for $RV_0$ and $\Delta
RV$ are respectively $-5$ and 20 km/s, the latter being
chosen to be rather inclusive}
since additional indicators are also
used for final membership assessment\footnote{ The value of mean radial
velocity $RV_0$ is in good agreement with that of the local molecular
gas, see e.g.\ Rebolledo \e (2016, their Fig.5), with heliocentric
$RV_0 = -5$~km/s corresponding to $V_{LSR}=-13.8$~km/s.}. According to the
above stated general rule, inside the dashed RV limits only one
membership criterion (the RV criterion being fulfilled) is sufficient to
consider a star as a member; outside of them, at least two indicators
are needed. The number of stars shown in the total histogram is 755;
scaled to this number, we expect 9 spurious optical-X-ray matches to be
included in these histograms, distributed uniformly {\em per star} (not
{\em per RV interval}): therefore, most of the spurious X-ray matches will lie
inside the dashed lines, where most stars are found. This justifies in
part the several bins containing X-ray detections in excess of lithium-rich
stars.
{ For the same reason, we expect 6-7 contaminants among candidate
members found only from RV and X-ray criteria.}

A handful of stars satisfying the
``general'' membership criteria above are found in the CMD very near the ZAMS
at the cluster distance (or anyway at apparent ages $>20$Myr according to
Siess \e 2000 isochrones), even after accounting for reddening as in the
following sections. Upon individual examination, only two\footnote{These
are the candidate HAeBe star [HSB2012] 2504, and the star [HSB2012]
2692, showing strong lithium, IR excess in the 2MASS bands, and
association with a YSO.}
of these 17 stars
showed strong enough indications of membership to be retained in our
list, while the others, mostly low-SNR spectra with poorly constrained
parameters, were removed. Of these latter, six were
simultaneously RV- and X-ray members, in very good agreement with their
expected number.

{ Among our targets, additional candidate CTTS members were searched, but
none found, using other catalogues: the VPHAS$+$ data, in
particular using the $(r-i,r-H\alpha)$ diagram locus as in Kalari \e
(2015); the 2MASS data; and finally the Spitzer data from Broos \e
(2011a; table~6), considering
disk-bearing stars with colors $[3.6-4.5]>0.2$ and $[5.8-8.0]>0.2$.
To this stage, the number of low-mass candidate Carina members (colder than
$\sim 8000$~K) found is 303, of which approximately 150 candidate CTTS.
}

We find also many stars with \teff\ above 6500~K, which already at this
young age have no traces of lithium, and are sometimes rotating so fast
that an accurate derivation of RV might not be possible. Moreover, it is known
that A-type stars are not strong coronal X-ray emitters (e.g., Schmitt 1997), so
that our member list is least reliable for stars earlier than type F
(approximately $T_{eff} \geq 7000$~K).
Early-A and B-type stars, of which we find several tens, are instead
rare in the field, so that most of them can be safely considered as
Carina members (adding to the 303 low-mass candidates),
even though their RVs might sometimes not be accurately
determined because of fast rotation.

\subsection{A young field-star population}
\label{young}

\begin{figure}
\resizebox{\hsize}{!}{
\includegraphics[bb=5 10 485 475]{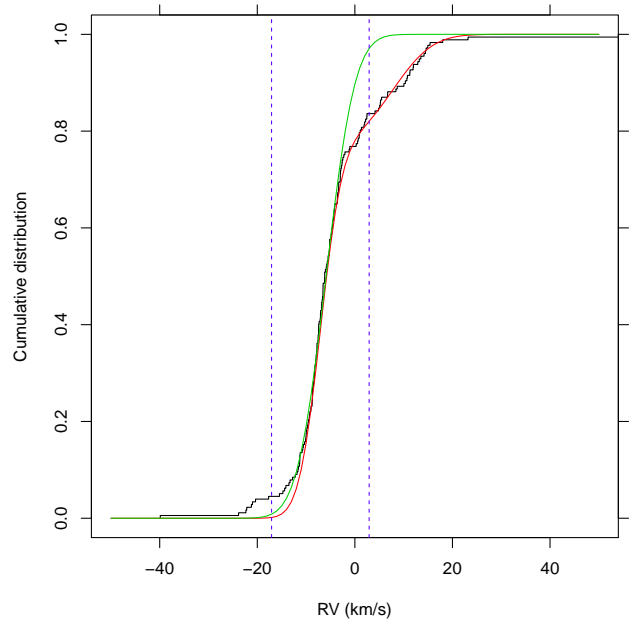}}
\caption{Cumulative RV distribution for late-type members with
$T_{eff}<7000$~K and $v \sin i<50$~km/s (solid line). Green: cumulative
single-Gaussian distribution; red: two-Gaussian distribution. The blue
vertical dashed lines indicate $\pm 7$~km/s from median RV, as a reference.
\label{rv-cumfun-fit}}
\end{figure}

\begin{figure}
\resizebox{\hsize}{!}{
\includegraphics[bb=5 10 485 475]{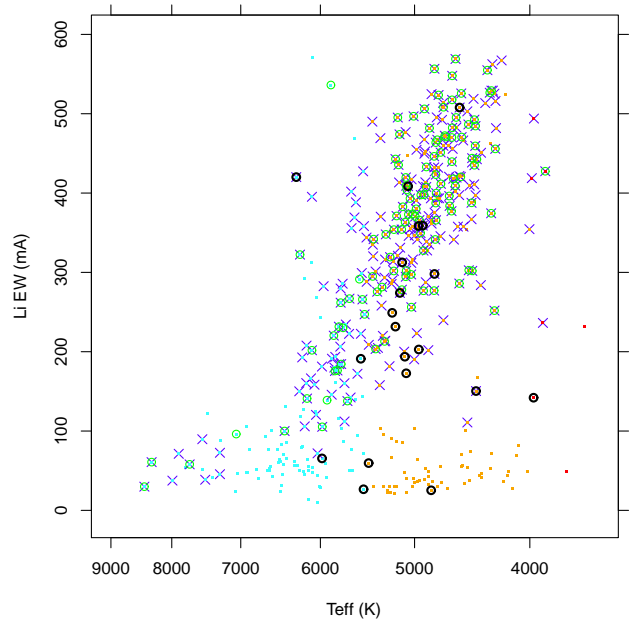}}
\caption{Diagram of lithium EW vs.\ \teff.
Color-coding of small
dots corresponds roughly to: solar-mass and earlier-type stars (cyan),
late-G to mid-K stars (orange), and M-type stars (red). Probable members
are indicated with blue crosses.
Green circles indicate (CTTS or Herbig Ae/Be) stars with wide \ha\ emission.
The bigger black circles indicate RV10 stars.
\label{teff-li}}
\end{figure}

A more detailed examination of candidate
members is done from their cumulative RV distribution, as
shown in Figure~\ref{rv-cumfun-fit}. The black curve is the RV
cumulative distribution for all single member stars with $T_{eff}<7000$~K,
$v \sin i<50$~km/s, and SNR$>15$ (177 stars). A cumulative Gaussian
distribution fitted to the median RV range between $(-20,40)$~km/s
is shown with the green curve (maximum-likelihood parameters
$<RV>=-5.9$~km/s, $\sigma(RV)=4.66$~km/s). However, an obvious
asymmetry is present between the two tails of the Gaussian
distribution, with an excess of stars at $RV>0$~km/s.
Unrecognized binaries are only expected to produce symmetric tails.
We inspected all spectra of stars with $RV>0$ to check that no RV
determination errors were present, to a level which might justify the
observed discrepancy.
The red curve
shows therefore the cumulative distribution corresponding to two
superimposed Gaussians, with maximum-likelihood parameters:
$<RV_1>=-7.07$~km/s,
$\sigma(RV)_1=3.48$~km/s; and $<RV_2>=7.81$~km/s, $\sigma(RV)_2=6.37$~km/s.
The main Gaussian distribution accounts for 77\% of stars in the
subsample considered here.
The still mis-fit tail at $RV \sim -20$~km/s may be attributed to
either binaries or an additional small population, of 6-7 stars, which
was not examined further.
The $\sim 40$ stars in the positive-velocity Gaussian are too many to
be explained by spurious X-ray matches, whose number was quantified
above: therefore, they are likely genuine X-ray bright and/or lithium
rich stars, or young stars in general. Their discrepant RV distribution
with respect to the main Carina population does not entitle us to
consider them as Carina members. Based on their approximate RV we will refer
to these stars as ``RV10 stars''. There are 29 RV10 stars
with $RV>3$~km/s, $T_{eff}<7000$~K, $v \sin i<50$~km/s,
and SNR$>15$.

\begin{figure*}
\sidecaption
\includegraphics[width=12cm]{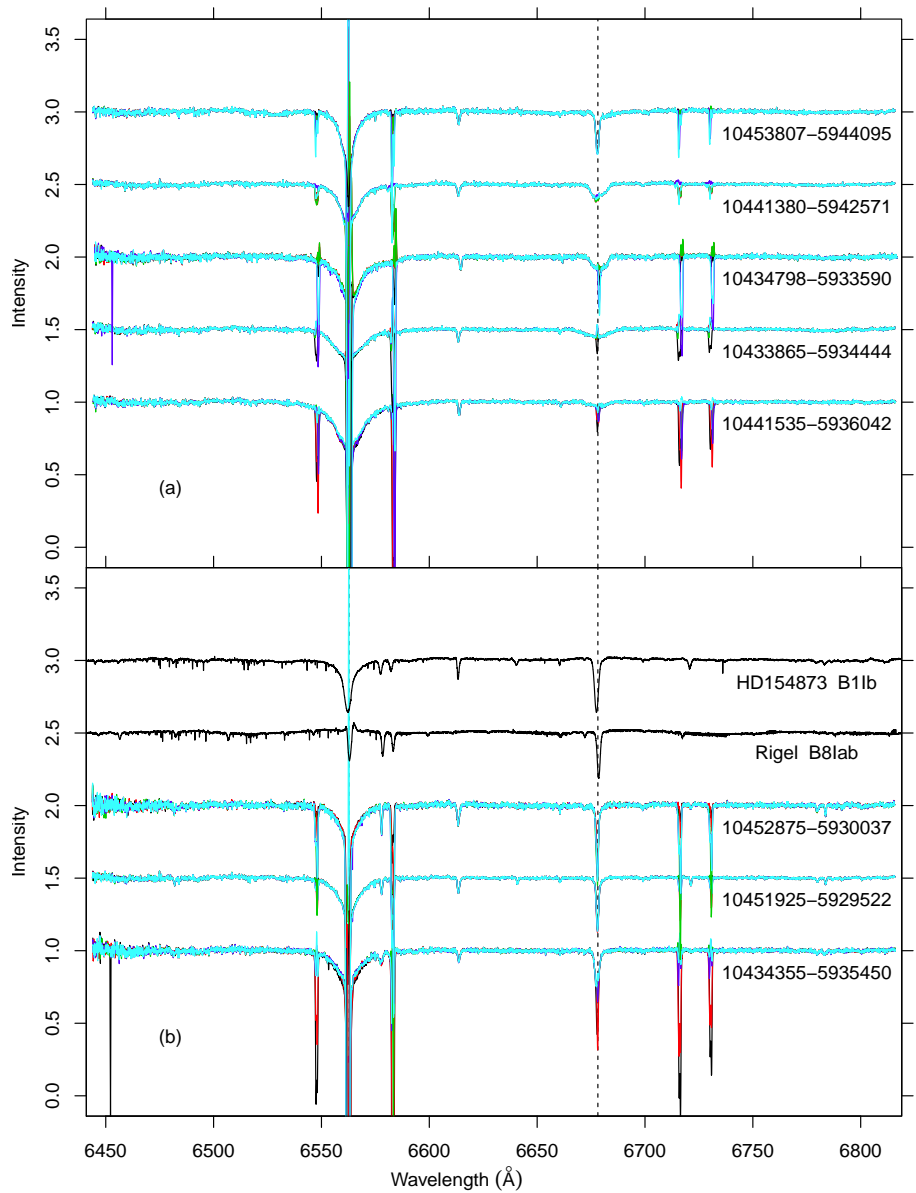}
\caption{
($a)$, top:
Example spectra of B stars found among our targets.
For each star, five spectra are overplotted with different colors,
corresponding to five distinct options regarding nebular subtraction.
The vertical dashed black line indicates the He~I 6678\AA\ line,
distinctive of B stars, but also coincident with a strong, narrow nebular line.
Spectra of different stars are offset by 0.5 for clarity.
The narrow absorption feature near 6613\AA\ is a well-known DIB, while a
weaker one is sometimes visible near 6660\AA.
$(b)$, bottom:
Spectra of sample B stars (lower three spectra) showing absorption in
the C~II doublet near 6580\AA\ (the redder line in the doublet coincides
with the nebular [N II] 6584\AA\ line), which is typical of known B bright
giants/supergiants (upper two spectra, from UVES/POP).
The vertical dashed lines indicate \ha\ and He~I 6678\AA.
\label{atlas-Bstars-2}}
\end{figure*}

We investigate the nature
of the RV10 group by means of the \teff-lithium EW diagram
(Figure~\ref{teff-li}): the RV10 stars account for most of the low-lithium
candidate members; however, more than one-half of the RV10 stars have
Li~EW $>150$~m\AA, and follow the same relation with \teff\ as Carina
members, to which are apparently coeval.
All RV10 stars with Li~EW $>250$~m\AA\ have extinction
(as determined in Section~\ref{reddening})
higher than the foreground $A_V^{fg}$ towards Carina (Hur \e 2012).
Nearly one-half of the RV10 stars
have instead low extinction, suggesting that they are foreground stars;
at the same time, many of these latter stars have significant Li EWs,
indicating they are young stars
(say $<100$~Myr), in agreement with their frequent X-ray detection
(18/29 stars). The spatial distribution of RV10 stars does not show
clustering, and their location on the optical CMD often does not overlap
the main cluster locus. We conclude that the RV10 are dominated by a group
of (17) foreground young stars, unrelated to the Carina cluster, probably
spanning a significant age range, and characterized by a common
dynamics. Also their $v \sin i$ distribution is typical of young low-mass stars.
We therefore speculate that they are a remnant from a now dissolved
cluster, closer to us than the Carina nebula, in the same sky direction.

The 12 RV10 stars with $A_V > A_V^{fg}$ are instead probable Carina
low-mass members, whose net number becomes 286.
Among them we find 8 SB2 binaries: this appears as a low percentage compared to
other studies (e.g.\ Mathieu 1994); however, fast
rotation, frequently found among Carina members (see
Section~\ref{rotation}) may render double-lined systems more difficult
to detect, so we cannot draw firm conclusions on this subject for Carina
stars. We do not consider these SB2 stars in the rest of the paper.
The identifications and properties of Carina members are listed in
Tables~\ref{members-ident} and~\ref{members-params},
while those of the RV10 group are in Tables~\ref{rv10-ident}
and~\ref{rv10-params}.

\begin{longtab}
\begin{longtable}{ccccccccc}
\caption{\label{members-ident} Identifications for late-type Carina members.
Column HSB2012 reports the identifier in Hur \e (2012); column
CXOGNC is the identifier in Broos \e (2011a); column ADM2008 is the
identifier in Albacete-Colombo \e (2008); column YSO is the identifier
in Zeidler \e (2016).
$V$ magnitudes, $B-V$ and $V-I$ colors are from Hur \e (2012).
}\\
\hline\hline
HSB2012 & CXOGNC & ADM2008 & YSO & RA & Dec & $V$ & $B-V$ & $V-I$ \\
 & & & & (J2000) & (J2000) & & & \\
\hline
\endfirsthead
\caption{continued.}\\
\hline\hline
HSB2012 & CXOGNC & ADM2008 & YSO & RA & Dec & $V$ & $B-V$ & $V-I$ \\
 & & & & (J2000) & (J2000) & & & \\
\hline
\endhead
\hline
\endfoot
796 & ... & ... & ... & 10:43:31.96 & -59:33:56.4 & 16.36 & 0.93 & 1.15 \\
800 &  104332.23-593511.3 & ... & ... & 10:43:32.25 & -59:35:11.5 & 16.20 & 0.80 & 1.11 \\
908 &  104336.77-593315.2 & ... & ... & 10:43:36.77 & -59:33:15.2 & 14.72 & 0.68 & 1.17 \\
907 & ... & ... & ... & 10:43:36.77 & -59:36:48.2 & 17.12 & 1.10 & 1.19 \\
954 &  104338.76-593301.9 & ... & ... & 10:43:38.80 & -59:33:01.8 & 16.21 & 1.08 & 1.50 \\
953 &  104338.81-593423.7 & ... & ... & 10:43:38.80 & -59:34:23.8 & 18.19 & ... & 2.33 \\
995 &  104340.13-593356.0 & ... & 104340.14-593356.3 & 10:43:40.13 & -59:33:56.1 & 18.32 & ... & 2.09 \\
998 &  104340.19-593449.5 & ... & ... & 10:43:40.18 & -59:34:49.6 & 17.04 & 1.55 & 2.13 \\
\end{longtable}
\end{longtab}

\begin{longtab}
\begin{longtable}{crr@{$\pm$}rr@{$\pm$}rc@{$\pm$}ccr@{$\pm$}rr@{$\pm$}rccc}
\caption{\label{members-params} Parameters of late-type Carina members.
Column HSB2012 reports the identifier in Hur \e (2012).
SNR is the spectrum median signal/noise ratio.
RV is the heliocentric radial velocity.
Errors on \teff\ are statistical only. The gravity index $\gamma$ is
defined in Damiani \e (2014).
Column ``log age'' is the
photometric age. Column \ha\ reports ``Y'' if an excess \ha\ emission is
detected, either from accretion (CTTS) or chromospheric, distinct from
the nebular \ha\ emission. Column ``SB2'' is ``Y'' for SB2 binaries.
}\\
\hline\hline
HSB2012 & SNR & \multicolumn{2}{c}{RV} & \multicolumn{2}{c}{$v \sin i$} &
\multicolumn{2}{c}{\teff} & $A_V$ & \multicolumn{2}{c}{Li EW} &
\multicolumn{2}{c}{$\gamma$} & log age &
\ha\ & SB2 \\
 & & \multicolumn{2}{c}{(km/s)} & \multicolumn{2}{c}{(km/s)} &
\multicolumn{2}{c}{(K)} & & \multicolumn{2}{c}{(m\AA)} &
\multicolumn{2}{c}{} & (yr) & \\
\hline
\endfirsthead
\caption{continued.}\\
\hline\hline
HSB2012 & SNR & \multicolumn{2}{c}{RV} & \multicolumn{2}{c}{$v \sin i$} &
\multicolumn{2}{c}{\teff} & $A_V$ & \multicolumn{2}{c}{Li EW} &
\multicolumn{2}{c}{$\gamma$} & log age &
\ha\ & SB2 \\
 & & \multicolumn{2}{c}{(km/s)} & \multicolumn{2}{c}{(km/s)} &
\multicolumn{2}{c}{(K)} & & \multicolumn{2}{c}{(m\AA)} &
\multicolumn{2}{c}{} & (yr) & \\
\hline
\endhead
\hline
\endfoot
796 &  17.0 &   2.5 &  2.5 &  20.3 &  3.3 & 5694 &  218 & 0.75 & 137.8 & 12.0 & 1.01 & 0.015 & ... & Y & N \\
800 &  43.1 &  -0.7 &  6.2 &  74.2 &  8.1 & 5978 &   95 & 0.81 & 105.3 &  6.4 & 1.009 & 0.006 & ... & Y & N \\
908 &  62.5 & -11.5 &  7.4 &  30.0 & 12.7 & 7541 &   92 & 1.77 &  89.7 &  3.9 & 0.999 & 0.004 & ... & N & N \\
907 &  28.5 &  -4.9 &  0.6 &  13.6 &  1.7 & 4869 &   95 & 0.31 & 202.0 &  6.6 & 0.969 & 0.009 & 6.98 & N & N \\
954 &  49.6 &  -6.5 &  8.0 &  69.0 &  2.7 & 5727 &   67 & 1.52 & 160.3 &  6.3 & 1.01 & 0.005 & ... & N & N \\
953 &  25.9 &  -3.6 &  1.4 &  24.8 &  1.5 & 4865 &  104 & 2.78 & 371.0 &  8.3 & 1.022 & 0.01 & 6.30 & N & N \\
995 &  16.6 &  -6.8 &  2.0 &  13.0 &  1.6 & 5019 &  172 & 2.44 & 373.8 & 11.0 & 0.999 & 0.016 & ... & Y & N \\
998 &  47.0 &  -9.3 &  7.0 &  92.4 &  1.4 & 4942 &   60 & 2.44 & 349.6 &  7.4 & 1.011 & 0.006 & 6.12 & Y & N \\
\end{longtable}
\end{longtab}

\begin{table*}
\centering
\caption{\label{rv10-ident} Identifications of RV10 stars.
Columns as in Table~\ref{members-ident}.
}
\begin{tabular}{ccccccccc}
\hline\hline
HSB2012 & CXOGNC & ADM2008 & YSO & RA & Dec & $V$ & $B-V$ & $V-I$ \\
 & & & & (J2000) & (J2000) & & & \\
\hline
394 &  104314.98-593341.3 & ... & ... & 10:43:14.99 & -59:33:41.3 & 14.67 & 0.94 & 1.09 \\
1262 &  104348.74-593430.4 & ... & ... & 10:43:48.59 & -59:34:31.2 & 17.72 & 1.76 & 2.03 \\
1788 &  104405.37-593043.5 & ... & ... & 10:44:05.42 & -59:30:43.5 & 14.53 & 0.90 & 1.03 \\
2169 & ... & ... & ... & 10:44:18.67 & -59:48:05.4 & 16.09 & 0.96 & 1.02 \\
2270 & ... & ... & ... & 10:44:22.78 & -59:38:52.5 & 18.44 & ... & 1.75 \\
2296 &  104423.64-593941.8 & 146 & ... & 10:44:23.56 & -59:39:41.4 & 16.34 & 0.91 & 1.34 \\
2561 & ... & ... & ... & 10:44:32.97 & -59:38:05.9 & 17.38 & 0.94 & 1.22 \\
2797 & ... & ... & ... & 10:44:42.77 & -59:38:01.4 & 17.03 & 1.14 & 1.14 \\
2959 &  104450.11-593429.3 & ... & ... & 10:44:49.99 & -59:34:29.5 & 16.75 & 1.32 & 1.72 \\
3030 &  104452.43-594130.1 & 469 & ... & 10:44:52.40 & -59:41:30.1 & 16.66 & 0.94 & 1.15 \\
3167 &  104457.05-593826.5 & 534 & ... & 10:44:57.06 & -59:38:26.9 & 13.70 & 1.01 & 1.19 \\
3229 & ... & ... & ... & 10:44:58.90 & -59:40:08.4 & 15.66 & 1.18 & 1.27 \\
3290 &  104501.02-594515.5 & 600 & ... & 10:45:01.03 & -59:45:15.6 & 14.54 & 0.91 & 1.06 \\
3688 & ... & ... & ... & 10:45:15.44 & -59:35:21.4 & 17.29 & 1.48 & 1.81 \\
3844 &  104521.23-593258.2 & ... & ... & 10:45:21.20 & -59:32:58.4 & 18.06 & 1.26 & 1.63 \\
3926 &  104524.13-593131.1 & ... & ... & 10:45:24.14 & -59:31:31.6 & 14.20 & 0.63 & 0.71 \\
4310 &  104541.60-593820.2 & 1014 & ... & 10:45:41.60 & -59:38:20.2 & 14.46 & 0.87 & 1.00 \\
\hline
\end{tabular}
\end{table*}

\begin{table*}
\centering
\caption{\label{rv10-params} Parameters of RV10 stars.
Columns as in Table~\ref{members-params}.
}
\begin{tabular}{crr@{$\pm$}rr@{$\pm$}rc@{$\pm$}ccr@{$\pm$}rr@{$\pm$}rccc}
\hline\hline
HSB2012 & SNR & \multicolumn{2}{c}{RV} & \multicolumn{2}{c}{$v \sin i$} &
\multicolumn{2}{c}{\teff} & $A_V$ & \multicolumn{2}{c}{Li EW} &
\multicolumn{2}{c}{$\gamma$} & log age &
\ha\ & SB2 \\
 & & \multicolumn{2}{c}{(km/s)} & \multicolumn{2}{c}{(km/s)} &
\multicolumn{2}{c}{(K)} & & \multicolumn{2}{c}{(m\AA)} &
\multicolumn{2}{c}{} & (yr) & \\
\hline
394 &  31.1 & 10.3 & 1.1 & 24.0 &  0.5 & 5189 & 105 & 0.39 & 231.6 & 7.2 & 0.984 & 0.009 & ... & N & N \\
1262 &  30.6 & 13.2 & 1.6 & 10.9 &  2.5 & 3830 &  17 & 0.42 & ... & ... & 0.862 & 0.008 & ... & Y & N \\
1788 &  84.4 & 23.2 & 0.4 & 12.0 &  0.5 & 5082 &  38 & 0.20 & 172.8 & 3.1 & 0.984 & 0.003 & ... & N & N \\
2169 &  40.1 & 14.2 & 0.6 & 13.9 &  1.6 & 5095 &  76 & 0.19 & 193.7 & 5.2 & 0.985 & 0.007 & ... & N & N \\
2270 &  17.4 &  4.8 & 2.5 & 17.9 & 10.0 & 3816 &  25 & 0.00 & ... & ... & 0.847 & 0.015 & ... & Y & N \\
2296 &  59.0 & 15.4 & 2.2 & 18.7 &  1.5 & 5548 &  66 & 1.10 & 191.1 & 5.3 & 1.012 & 0.004 & ... & N & N \\
\hline
\end{tabular}
\end{table*}

\begin{table*}
\centering
\caption{\label{giants-ident} Identifications of lithium-rich field giants.
Columns as in Table~\ref{members-ident}.
}
\begin{tabular}{ccccccccc}
\hline\hline
HSB2012 & CXOGNC & ADM2008 & YSO & RA & Dec & $V$ & $B-V$ & $V-I$ \\
 & & & & (J2000) & (J2000) & & & \\
\hline
3579 & ... & ... & ... & 10:45:11.09 & -59:40:07.8 & 15.85 & 1.96 & 5.05 \\
2595 & ... & ... & ... & 10:44:34.42 & -59:42:57.8 & 15.87 & 2.49 & 3.20 \\
3504 & ... & ... & ... & 10:45:08.53 & -59:43:30.2 & 18.10 & ... & 3.25 \\
\hline
\end{tabular}
\end{table*}

\begin{table*}
\centering
\caption{\label{giants-params} Parameters of lithium-rich field giants.
Columns as in Table~\ref{members-params}.
}
\begin{tabular}{crr@{$\pm$}rr@{$\pm$}rc@{$\pm$}ccr@{$\pm$}rr@{$\pm$}rccc}
\hline\hline
HSB2012 & SNR & \multicolumn{2}{c}{RV} & \multicolumn{2}{c}{$v \sin i$} &
\multicolumn{2}{c}{\teff} & $A_V$ & \multicolumn{2}{c}{Li EW} &
\multicolumn{2}{c}{$\gamma$} & log age &
\ha\ & SB2 \\
 & & \multicolumn{2}{c}{(km/s)} & \multicolumn{2}{c}{(km/s)} &
\multicolumn{2}{c}{(K)} & & \multicolumn{2}{c}{(m\AA)} &
\multicolumn{2}{c}{} & (yr) & \\
\hline
3579 & 133.3 & 22.1 & 2.3 & 24.7 & 4.6 & 2819 &  24 & ... & 487.5 & 2.1 & 1.019 & 0.002 & ... & N & N \\
2595 &  96.9 & 17.7 & 1.4 & 18.9 & 1.6 & 3598 & 150 & 2.20 & 231.3 & 2.1 & 1.061 & 0.003 & ... & N & N \\
3504 &  32.0 & 43.1 & 0.7 &  7.5 & 1.7 & 4424 &  69 & 4.24 & 167.5 & 5.9 & 1.051 & 0.008 & ... & N & N \\
\hline
\end{tabular}
\end{table*}

\begin{longtab}
\begin{landscape}
\begin{longtable}{llcccccll}
\caption{\label{bstars} Early-type stars.
Column ``SpT'' reports the spectral type listed in SIMBAD.
Column ``Notes'' reports our classification, in particular: ``B SG"
indicates candidate B supergiants; ``HAeBe'' indicates Herbig Ae/Be
stars; ``near A0'' indicates stars of late-B/early-A type.
}\\
\hline\hline
2MASS Id & SIMBAD Id & RA & Dec & $V$ & $B-V$ & $V-I$ & SpT & Notes \\
\hline
\endfirsthead
\caption{continued.}\\
\hline\hline
2MASS Id & SIMBAD Id & RA & Dec & $V$ & $B-V$ & $V-I$ & SpT & Notes \\
\hline
\endhead
\hline
\endfoot
10430716-5931209 & [HSB2012] 230 & 10:43:07.16 & -59:31:20.9 & 15.06 & 1.08 & 1.76 & ... & B type \\
10432086-5929595 & [HSB2012] 538 & 10:43:20.86 & -59:29:59.6 & 14.43 & 0.39 & 0.72 & ... & B type \\
10433865-5934444 & Cl Trumpler 14 43 & 10:43:38.66 & -59:34:44.5 & 12.17 & 0.42 & 0.71 & ... & B type \\
10434798-5933590 & Cl Trumpler 14 45 & 10:43:47.98 & -59:33:59.1 & 13.2 & 0.44 & 0.73 & ... & B type \\
10435224-5936585 & [HSB2012] 1385 & 10:43:52.24 & -59:36:58.5 & 12.24 & 0.4 & 0.7 & ... & B type \\
10435250-5942503 & [HSB2012] 1395 & 10:43:52.50 & -59:42:50.4 & 16.47 & 1.1 & 1.8 & ... & B type \\
10435723-5932411 & Cl* Trumpler 14 VBF 65 & 10:43:57.24 & -59:32:41.2 & 12.85 & 0.31 & 0.55 & ... & B type \\
... & [HSB2012] 1618 & 10:43:59.20 & -59:33:21.5 & 13.44 & 0.39 & 0.67 & ... & B type \\
\end{longtable}
\end{landscape}
\end{longtab}

Three stars with strong lithium lines but showing no other membership
indicator turned out to be non-member lithium-rich giants, upon
a detailed examination of their stellar parameters. Their
identifications and properties are
reported in Tables~\ref{giants-ident} and~\ref{giants-params}.

\section{Massive stars}
\label{ob}

Our observed sample selection was made with the aim of studying low-mass
stars in Carina, starting from A-type stars. However, a number
of earlier-type stars are serendipitously found among the spectra
studied here. Those O-type and B-type stars are intrinsically more
luminous than the rest of the sample,
but some were included here
because of their higher extinctions or, in some cases, because
they are located at significantly longer distances (the Carina Nebula is
close to the tangent of a spiral arm).
Although HR15n spectra are not ideal for a detailed
classification of hot stars, we attempt here to assign a qualitative
classification of these stars, whose importance in the study of the
Carina SFR cannot be neglected. The spectroscopic information is
complemented with a study using archival photometry.

\subsection{Spectroscopic analysis}

\begin{figure*}
\sidecaption
\includegraphics[width=12cm]{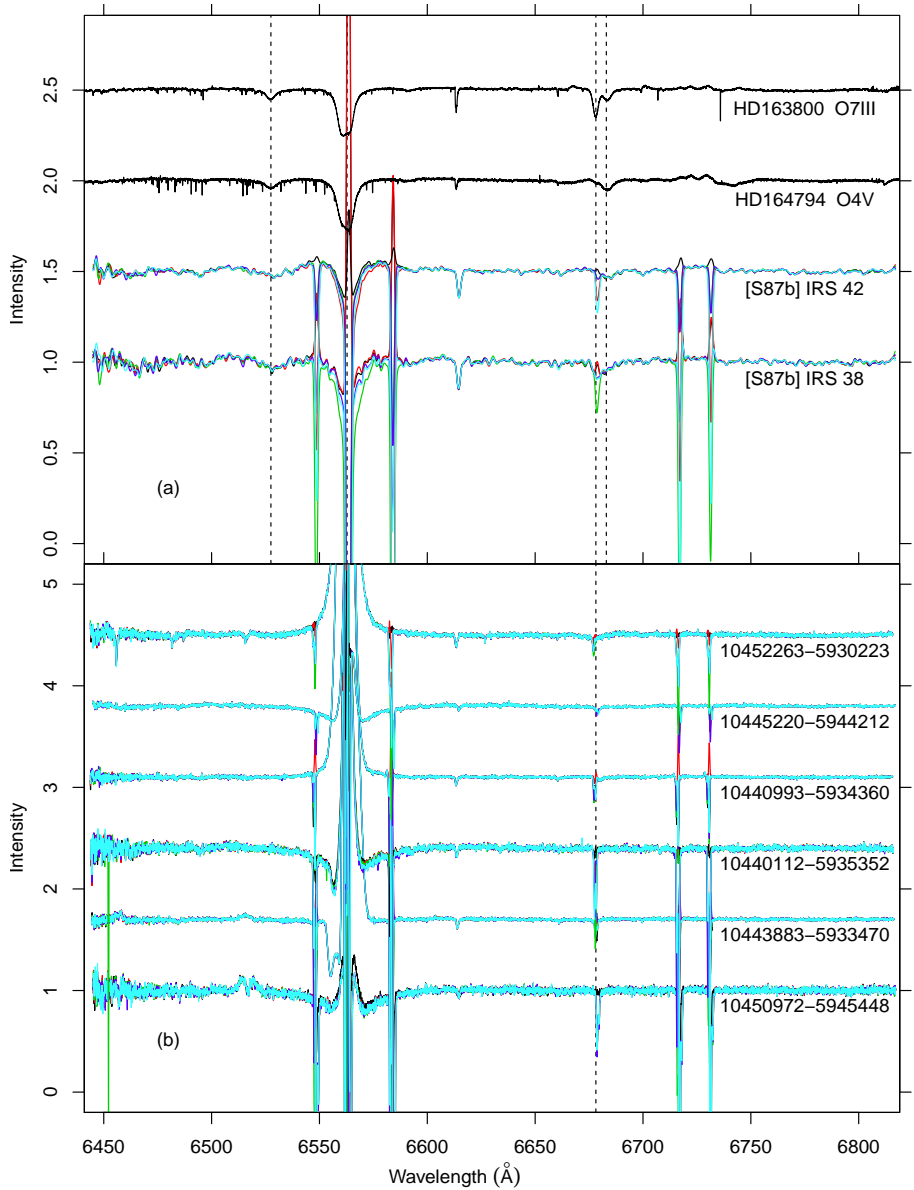}
\caption{
($a)$, top:
Spectra of a possible O star (SIMBAD identifier [S87b] IRS 38),
a probable O star ([S87b] IRS 42), and two UVES/POP spectra of known O
stars. The lower two spectra were slightly smoothed to improve their
visibility. Dashed vertical lines indicate \ha\ and He~I 6678\AA, as
above, plus the He~II lines at 6527 and 6684\AA, typical of O stars.
($b)$, bottom:
Example spectra of new candidate Herbig Ae/Be stars discovered
in this work. Note the consistently wide \ha\ emission, regardless of
the option chosen for nebular subtraction (color of plotted spectrum).
Some of the stars show also absorption/emission in the Fe~II lines at
6456.38, 6516.08\AA.
\label{atlas-Ostars-Herbig}}
\end{figure*}

As mentioned in Sect.~\ref{obs}, the HR15n wavelength range contains the
He~I 6678.15\AA\ line, which is a very useful diagnostics to select
B-type stars. Of course, care must be used since at nearly the same
wavelength a Fe~I line is found in cooler stars, so that establishing a
B-type spectrum also requires absence of metallic lines, and especially
the group of lines between 6490-6500\AA\ (Damiani \e 2014), strong
enough to remain recognizable even in fast rotators.
As explained in Sect.~\ref{obs} above (and see Paper~I for details),
\ha\ and He~I lines, which are the most intense lines in the HR15n range
for late-B stars, and the only detectable at low SNR, are both
coincident with nebular sky lines, whose contribution is often
impossible to subtract accurately. Therefore, classification of late-B
stars on the basis of the present data may only be approximate. In
practice it is often impossible to confirm or exclude the presence of a
weak He~I line in the spectrum of a fast-rotating star without metallic
lines, to discriminate between a late-B and an early-A type, and
therefore we will assign to these stars a generic "late-B/early-A" type.
In tens of other cases, He~I is safely detected, but again nebular lines
make the determination of the stellar He~I EW unreliable: these stars
were thus classified as "B-type" without better detail.
Examples of B-type stars in our dataset are shown in
Fig.~\ref{atlas-Bstars-2}-$a$.
In this Figure,
differences in the
adopted sky correction are evident only in the cores of nebular lines
\ha, [N II] 6548,6584\AA, He~I, and [S II] 6717,6731\AA.

Some of the B stars show a characteristic C~II doublet at 6578-6583\AA,
which is typical of mid-B stars, but is enhanced in bright giants or
supergiants (Grigsby \e 1992, Sigut 1996, McErlean \e 1999).
We therefore consider these stars as candidate B supergiants, although
some of them may be main-sequence B stars as well.
We show some of these spectra in Figure~\ref{atlas-Bstars-2}-$b$,
together with spectra of two known B supergiants (top two spectra) from
the UVES/POP dataset (Bagnulo \e 2003).

The Gaia-ESO spectra of two of our stars show features suggesting an O
stellar type: these are the He\,{\sc ii} lines at 6527~\AA\ and 6683~\AA. Their
spectra are shown in Figure~\ref{atlas-Ostars-Herbig}-$a$ (bottom two spectra),
compared with known O stars from UVES/POP.
The intensity ratio between He\,{\sc i}~6678~\AA\ and the neighboring
He\,{\sc ii}~6683~\AA\ line would be a sensitive measure of spectral type;
however, the nebular contamination at 6678~\AA\ is strong for both stars,
since they are highly reddened and faint, which prevents accurate derivation
of their spectral types.
Furthermore, a large fraction of OB stars are spectroscopic binaries
(Sota \e 2014)  
and an examination of these lines in
the sample covered by spectroscopic surveys such as OWN
(Barb\'a \e 2010),  IACOB   
(Sim\'on-D\'{\i}az \e 2011),   
or CAF\'E-BEANS
(Negueruela \e 2015)         
reveal a number of cases where a B+B SB2 (each star with a He\,{\sc
i}~6678~\AA\ line) masquerades as a single object with both He\,{\sc i}
and He\,{\sc ii} lines. Finally, He\,{\sc ii}~6527~\AA\ is close to a
broad DIB.
Therefore, their identification as O stars based on their
spectra is tentative at this stage.
Both stars have already appeared in the literature on Carina stars,
first as bright IR sources in Smith (1987 - the corresponding
designation is reported in Fig.~\ref{atlas-Ostars-Herbig}-$a$), then as
candidate massive stars from both X-ray and IR observations (Sanchawala \e 2007,
Povich \e 2011a); no optical spectra were however published for these
stars before.

Finally, 17 stars are good or possible candidates as Herbig Ae/Be stars,
having an A/B type and definite (regardless of sky-subtraction option
chosen) wide emission at \ha. We show some examples of these stars'
spectra in Figure~\ref{atlas-Ostars-Herbig}-$b$.
Some of the spectra show also absorption or emission in the Fe~II lines at
6456.38, 6516.08\AA, indicative of hot gas. While many of these stars'
spectra show clearly wide \ha\ absorption wings, underlying the
circumstellar \ha\ emission, in some cases the absorption component is
overwhelmed by the wide emission component: in these cases the HAeBe
nature is suggested by either a He~I line, by the Fe~II lines, or by the
optical photometry, better consistent with an A-B type than with a
lower-mass CTTS.

\begin{figure}
\resizebox{\hsize}{!}{
\includegraphics[bb=5 10 485 475]{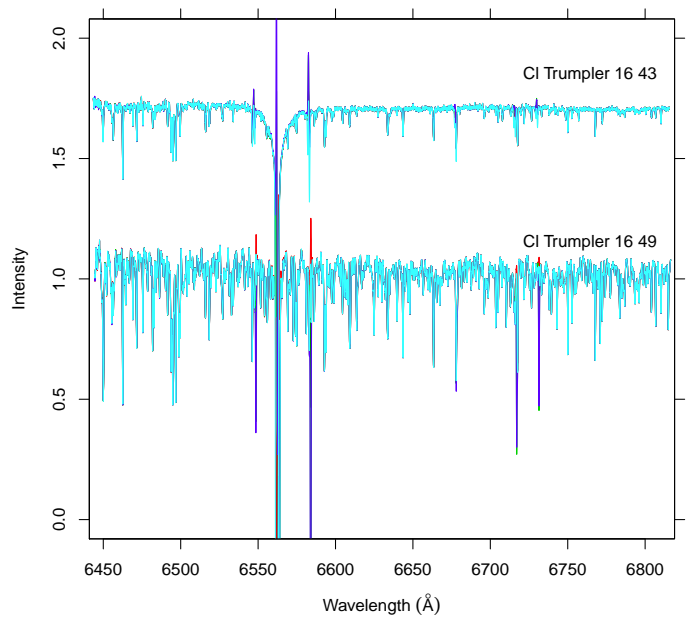}
}
\caption{Spectra of two stars mis-classified in the SIMBAD database as B
stars, while they are respectively a K giant (lower spectrum),
and a late-A or F star (upper spectrum).
Spectra are labeled with their SIMBAD identifier.
Color coding as in Figure~\ref{atlas-Bstars-2}.
\label{atlas-bad-simbad}}
\end{figure}

All massive stars found in our spectral dataset are listed in
Table~\ref{bstars}.
The 'SIMBAD Id' column shows that all of
these stars have also appeared in the SIMBAD database; however,
for only 12 of them a B spectral
type was already determined: therefore, the large majority of our
early-type stars are new entries in the massive-star population of
Carina, including all HAeBe stars.
Incidentally, our complete spectral sample also includes two stars for which
SIMBAD lists a B spectral type (with no reference given), while our
spectra show definitely later (F and K) spectral types, as shown in
Figure~\ref{atlas-bad-simbad}.

We have also matched our stellar sample with the lists of candidate
massive stars from Sanchawala \e (2007) and Povich \e (2011a).
Of the seven matches with the Sanchawala \e (2007) candidates, we
confirm a type earlier than early-A for 4 (at most 5) stars;
only four of our stars match the Povich \e (2011a) candidates, and all of
them are confirmed as early-type stars.
The combined list contains 11 massive star candidates, of which only 2-3
are not confirmed. Again, these are small numbers with respect to the
number of new early-type stars found for the first time in our Gaia-ESO
spectra (110 stars in Table~\ref{bstars}), in a spatial region much
smaller than the whole Carina region covered by the Povich \e (2011a)
study.

\subsection{Photometric analysis with CHORIZOS}

Some of the massive stars identified in this section are potentially
interesting because their colors suggest large extinctions
($A_V\sim 5$ or higher) which are rarely found for objects immersed in
H\,{\sc ii} regions.
For that reason we selected 10 stars
from Table~\ref{bstars} that are potential candidates for being highly
extinguished.
For seven of the stars $UBVIJHK$ photometry is available from Hur \e
(2012) and 2MASS (Table~\ref{ubvijhk}) while
for the remaining three only $BVIJHK$ photometry exists
(Table~\ref{bvijhk}).

We processed the photometry described above using the CHORIZOS code
(Ma{\'{\i}}z~Apell{\'a}niz 2014)  
in order to measure the extinction experienced by the
stars and, to the degree in which it is possible, their effective temperature $T_{\rm eff}$ and logarithmic distance $\log d$. The procedure 
followed is similar to the one used in other papers such as
Ma{\'{\i}}z~Apell{\'a}niz \e (2015).   
{
With the purpose of obtaining independent assessments of these stars'
properties, the results from the spectroscopic analysis above were not
used as inputs here.
}
We start by detailing the $UBVIJHK$ photometry case:

\begin{itemize}
 \item We used the Milky Way grid of
Ma{\'{\i}}z~Apell{\'a}niz (2013$a$),  
grid parameters are effective temperature ($T_{\rm eff}$) and photometric
luminosity class (LC). The latter quantity is defined in an analogous
way to the spectroscopic equivalent, but
{ maps discrete classes I-V to a continuous variable in the range
[0-5.5].}
For objects with 
$T_{\rm eff} > 15$ kK the spectral energy distributions (SEDs) are TLUSTY
(Lanz and Hubeny 2003). 
 \item The extinction laws were those of
Ma{\'{\i}}z~Apell{\'a}niz \e (2014),  
which are a single-family parameter with the type of extinction defined by 
$R_{5495}$ { (monochromatic analogous of $R_V$, see
Ma{\'{\i}}z~Apell{\'a}niz 2013$b$). The amount of reddening
is parameterized by $E(4405-5495)$.}
 \item LC was fixed to 5.0 (``main sequence'') in all cases except for $[$HSB2012$]$~3994, for which it was fixed to 1.0 (``supergiant''). 
       $T_{\rm eff}$, $R_{5495}$, $E(4405-5495)$, and $\log d$ were left as free parameters. 
\end{itemize}

The $BVIJHK$ photometry case was similar but it required an additional step. The problem with lacking $U$-band photometry is that 
$T_{\rm eff}$ and $\log d$ become quasi-degenerate (even if LC is known). $R_{5495}$ and $E(4405-5495)$, however, are strongly constrained, as
the different combinations of valid $T_{\rm eff}$ and $\log d$ yield almost identical values of the two extinction parameters. Therefore, the
procedure we followed was to select a reasonable fixed value of $\log d$ and leave $T_{\rm eff}$, $R_{5495}$, and $E(4405-5495)$ as free
parameters.

The results of the CHORIZOS analysis are presented in Tables~\ref{ubvijhk}~and~\ref{bvijhk}.

\begin{figure}
 \centerline{\includegraphics[width=\linewidth]{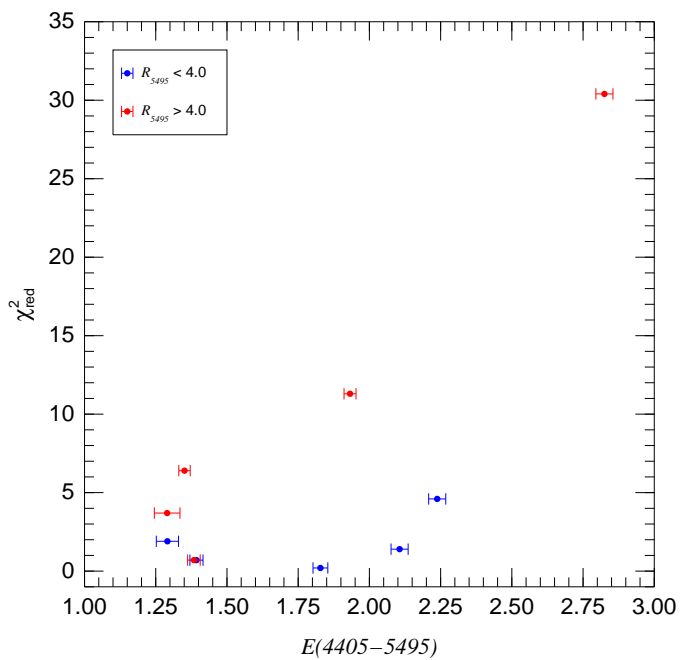}}
\caption{$\chi^2_{\rm red}$ vs.\ $E(4405-5495)$ for the ten highly extinguished stars analyzed with CHORIZOS.}
\label{ebvchir}
\end{figure}

\begin{itemize}
 \item We find values of $R_{5495}$ between $\approx$3.6 and $\approx$4.9, higher than the Galactic average but typical for an H\,{\sc ii} region.
 \item $E(4405-5495)$ ranges between $\approx$1.3 and $\approx$2.8, reflecting the known strong differential extinction. The three stars with
       no $U$-band photometry have the three highest values of $E(4405-5495)$, as expected.
 \item Typically, $A_V$ is determined with better precision than either $R_{5495}$ or $E(4405-5495)$ even though, to a first 
       approximation,
$A_V \approx R_{5495} \cdot E(4405-5495)$. This is not incorrect because the likelihood 
       in the $R_{5495}$-$E(4405-5495)$ plane is an elongated quasi-ellipsoid with negative slope i.e. $R_{5495}$ and $E(4405-5495)$ are 
       anticorrelated. This effect is commonly seen in fits to extinguished hot stars with CHORIZOS or similar codes.
 \item The most important result regarding the extinction is the location of four stars ($[$S87b$]$ IRS 41, $[$S87b$]$ IRS 42, 
       $[$HSB2012$]$ 1920, and $[$S87b$]$ IRS 38) in the $E(4405-5495)$-$R_{5495}$ plane. A comparison with Figure~2 of
Ma{\'{\i}}z~Apell{\'a}niz (2015) reveals  
       that they are in a region where no other known Galactic OB stars are located. That is because for low values of $E(4405-5495)$ it is 
       possible to find a large range of $R_{5495}$ values but for large reddenings $R_{5495}$ tends to be close to the average $\sim$3.1 value. 
       In other words, those four stars are exceptional in having large
values of both $E(4405-5495)$ and $R_{5495}$, { and their extinctions laws
deserve a more detailed study.}
 \item We plot in Fig.~\ref{ebvchir} the dependence of $\chi^2_{\rm red}$ of the CHORIZOS fit with reddening. For $R_{5495} < 4.0$, the fits are
       reasonably good even for values of $E(4405-5495) \sim 2.0$, indicating the validity of the extinction laws of
Ma{\'{\i}}z~Apell{\'a}niz \e (2014)   
        for low $R_{5495}$. For $R_{5495} > 4.0$, there are good fits for low values of the reddening but $\chi^2_{\rm red}$ grows increasingly large
       as $E(4405-5495)$ increases, indicating that the extinction laws of
Ma{\'{\i}}z~Apell{\'a}niz \e (2014)   
need to be modified for large values of $R_{5495}$.
       This is not surprising, given that those laws were derived using stars with lower reddenings and here we are extrapolating to 
       much larger values. 
 \item For two of the B-type stars (as determined from the spectroscopy) with $UBVIJHK$ photometry ($[$HSB2012$]$ 3017 and $[$HSB2012$]$ 3994) 
       the CHORIZOS-derived $T_{\rm eff}$ is consistent with them being mid-B stars. For the other five objects with $UBVIJHK$ photometry
       (from the spectroscopy, three B types, one B SG and one O type), the uncertainties in $T_{\rm eff}$ are larger and the values lean towards
       them being O stars. However, we should be cautious of that result, especially for the cases with high $\chi^2_{\rm red}$. They could be 
       late-O stars (with weak He\,{\sc ii}~6683~\AA) or the fitted values for $T_{\rm eff}$ could be biased by the extinction law
(Ma{\'{\i}}z~Apell{\'a}niz \e 2014).  
 \item $[$HSB2012$]$ 3994 has an exceptionally large fitted distance (beyond the expected extent of the Galactic disk). One possible solution
       is that it is not a supergiant but instead a lower luminosity star. Even then, it is likely to lie beyond the Carina Nebula.
 \item Three other stars ($[$HSB2012$]$ 230, $[$HSB2012$]$ 1395, and $[$HSB2012$]$ 2913) are also likely to lie beyond the Carina Nebula. On the
       other hand, the fitted $\log d$ for $[$HSB2012$]$ 3017, $[$S87b$]$ IRS 41, and $[$S87b$]$ IRS 42 are compatible with them belonging to the
       Carina Nebula Association. It should be pointed out that the latter list includes the two stars with the highest values of $R_{5495}$. That
       is an expected effect because for stars beyond the Carina Nebula one expects a larger contribution to extinction from the diffuse ISM and, 
       hence, a lower $R_{5495}$.
\end{itemize}

\begin{table*}
\centering
\caption{Results of the CHORIZOS fits for the seven highly extinguished stars with $UBVIJHK$ photometry.}
\label{ubvijhk}
\begin{tabular}{lr@{$\pm$}lcr@{$\pm$}lr@{$\pm$}lr@{$\pm$}lr@{$\pm$}lc}
\hline
Star               & \multicolumn{2}{c}{$T_{\rm eff}$} & LC  & \multicolumn{2}{c}{$R_{5495}$} & \multicolumn{2}{c}{$E(4405-5495)$} & \multicolumn{2}{c}{$\log d$} & \multicolumn{2}{c}{$A_V$} & $\chi^2_{\rm red}$ \\
                   & \multicolumn{2}{c}{kK}            &     & \multicolumn{2}{c}{}           & \multicolumn{2}{c}{mag}            & \multicolumn{2}{c}{pc}       & \multicolumn{2}{c}{mag}   &                    \\
\hline
$[$HSB2012$]$ 3017 & 22.2 & 2.5                        & 5.0 & 3.62 & 0.08                    & 1.291 & 0.039                      & 3.50 & 0.07                  & 4.688 & 0.077             &  1.9               \\ 
$[$HSB2012$]$ 230  & 41.8 & 5.3                        & 5.0 & 3.95 & 0.07                    & 1.393 & 0.023                      & 3.88 & 0.11                  & 5.531 & 0.036             &  0.8               \\ 
$[$HSB2012$]$ 1395 & 42.3 & 5.2                        & 5.0 & 4.20 & 0.08                    & 1.384 & 0.022                      & 4.11 & 0.11                  & 5.831 & 0.040             &  0.7               \\ 
$[$HSB2012$]$ 2913 & 40.8 & 6.0                        & 5.0 & 3.69 & 0.05                    & 1.828 & 0.026                      & 3.85 & 0.12                  & 6.756 & 0.040             &  0.2               \\ 
$[$HSB2012$]$ 3994 & 18.5 & 3.2                        & 1.0 & 4.57 & 0.10                    & 1.290 & 0.045                      & 4.51 & 0.05                  & 5.902 & 0.114             &  3.7               \\ 
$[$S87b$]$ IRS 41  & 42.4 & 4.4                        & 5.0 & 4.92 & 0.09                    & 1.351 & 0.020                      & 3.37 & 0.09                  & 6.669 & 0.034             &  6.4               \\ 
$[$S87b$]$ IRS 42  & 42.0 & 4.2                        & 5.0 & 4.60 & 0.06                    & 1.932 & 0.021                      & 3.22 & 0.09                  & 8.860 & 0.033             & 11.3               \\ 
\hline
\end{tabular}
\end{table*}

\begin{table*}
\centering
\caption{Results of the CHORIZOS fits for the three highly extinguished stars with $BVIJHK$ photometry.}
\label{bvijhk}
\begin{tabular}{lr@{$\pm$}lr@{$\pm$}lr@{$\pm$}lc}
\hline
Star               & \multicolumn{2}{c}{$R_{5495}$} & \multicolumn{2}{c}{$E(4405-5495)$} & \multicolumn{2}{c}{$A_V$} & $\chi^2_{\rm red}$ \\
                   & \multicolumn{2}{c}{}           & \multicolumn{2}{c}{mag}            & \multicolumn{2}{c}{mag}   &                    \\
\hline
$[$HSB2012$]$ 3880 & 3.64 & 0.06                    & 2.106 & 0.030                      &  7.636 & 0.040            &  1.4               \\ 
$[$HSB2012$]$ 1920 & 3.93 & 0.06                    & 2.238 & 0.030                      &  8.757 & 0.026            &  4.6               \\ 
$[$S87b$]$ IRS 38  & 4.07 & 0.05                    & 2.825 & 0.030                      & 11.394 & 0.034            & 30.4               \\ 
\hline
\end{tabular}
\end{table*}

{
We should note that the uncertainties derived by CHORIZOS are relatively
small but not unrealistic because for over a decade we have worked to
eliminate the systematic errors that have plagued other works. That
includes recalculating the zero points for different photometric systems
(Ma{\'{\i}}z~Apell{\'a}niz 2005, 2006, 2007),  
complementing and testing different atmosphere grids
(Ma{\'{\i}}z~Apell{\'a}niz 2013a),  
deriving a new family of extinction laws that provides significant better
fits to photometric data than pre-existing ones
(Ma{\'{\i}}z~Apell{\'a}niz 2013b, Ma{\'{\i}}z~Apell{\'a}niz \e 2014a),
obtaining a library of
data on reference stars to serve as a test bed for all of the above
(Ma{\'{\i}}z~Apell{\'a}niz \e 2004, (Ma{\'{\i}}z~Apell{\'a}niz and Sota
2008, Ma{\'{\i}}z~Apell{\'a}niz \e 2011),
and integrating everything into the code. }

\section{Spatial groups}
\label{spatial}

\begin{figure}
\resizebox{\hsize}{!}{
\includegraphics[bb=5 10 485 475]{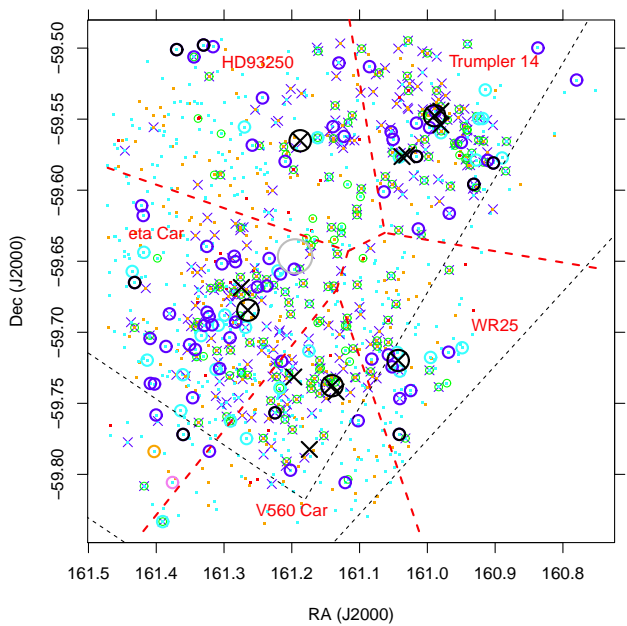}}
\caption{Spatial map of observed stars. The lower right region is empty
because of our sample selection choice.
Symbols as in Figure~\ref{teff-li}, with the addition of
big circles to indicate: B stars (blue),
late-B to early-A stars (cyan), candidate B supergiants (black), and
probable (orange) or possible (violet) O stars.
We also plot here O stars from Walborn (1973) as
black crosses. The most massive of them (surrounded by big black
circles) are taken as centers of the respective sub-regions, bounded by
red dashed lines and labeled in red after the central star names (except for
Trumpler~14, whose central star is the O2If*$+$O3.5 binary HD93129A).
Dashed black segments delimit the
dark V-shaped dust lanes in front of the Nebula.
The solid grey circle to the NW of $\eta$ Car indicates approximate
location and size of the Keyhole nebula (Smith and Brooks 2008).
\label{ra-dec-bstars}}
\end{figure}

\begin{figure}
\resizebox{\hsize}{!}{
\includegraphics[bb=5 10 485 475]{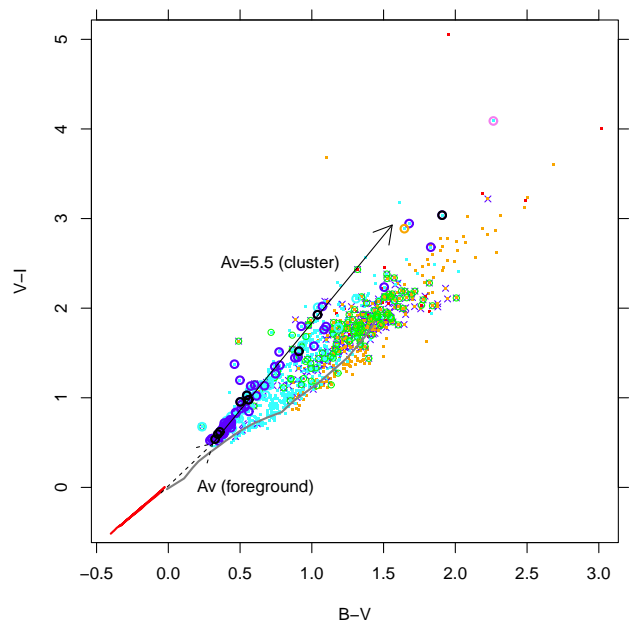}}
\caption{A $(B-V,V-I)$ color-color diagram of the observed sample.
Symbols as in Figs.~\ref{teff-li} and~\ref{ra-dec-bstars}.
The thick red line is the unreddened locus of stars earlier than A0,
while the thick grey line is the same for later-type stars. The
reddening law appropriate for cluster stars is illustrated by the solid
arrow, on top of the dashed arrow describing foreground absorption (and
reddening law). The length of the solid arrow indicates the reddening of
the new candidate O star 2MASS10453674-5947020 (star [S87b] IRS 42 in
SIMBAD).
\label{bv-vi-bstars}}
\end{figure}

\begin{figure*}
\centering
\includegraphics[bb=0 0 576 577,width=17cm,angle=0]{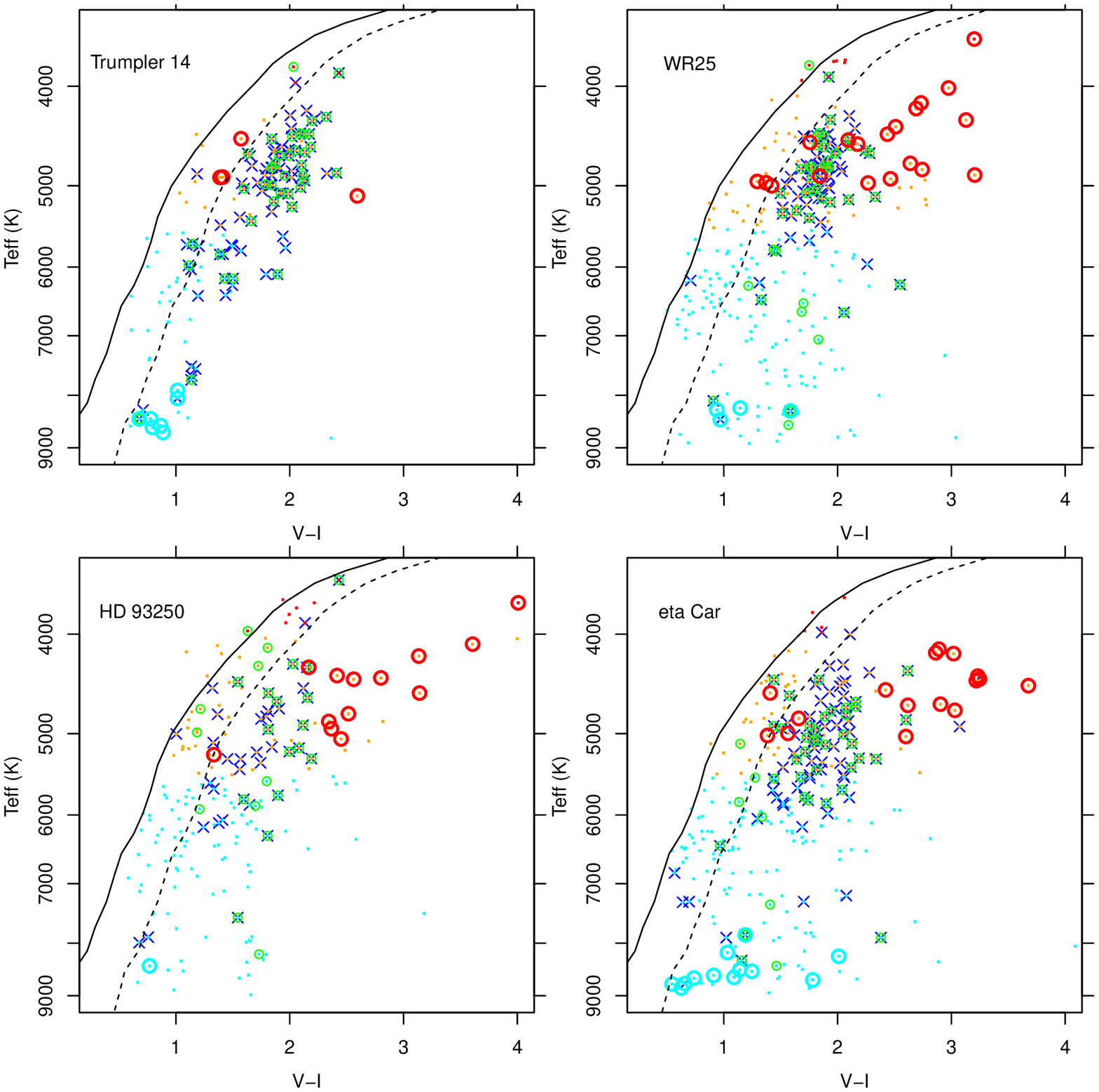}
\caption{Diagrams of \teff\ vs.\ color $V-I$, for four spatial
subregions as indicated. The KH95 \teff-color relation is shown for zero
reddening (solid) and for the cluster foreground reddening (dashed).
Symbols as in Figs.~\ref{teff-li} and~\ref{ra-dec-bstars}.
Big red circles indicate luminous low-gravity
giants, probably located behind the Nebula.
\label{teff-vi-groups}}
\end{figure*}

The Carina Nebula is known to host a morphologically complex stellar
population, distributed among approximately 20 sub-clusters and a sparse
population according to Feigelson \e (2011), of which Tr~14 and Tr~16
are only the most massive. Therefore, we define here several spatial
groupings into which we conveniently split our sample stars.
Among clusters defined in Feigelson \e (2011), numbers 1, 3-6, 9-12, 14
fall in the region studied here. This is however a too detailed
subdivision considering the number of optical spectra available here, so
we prefer a simpler categorization. In Figure~\ref{ra-dec-bstars} we
show the spatial distribution of all stars observed in Gaia-ESO, and
known massive stars as a reference.
The most massive stars are also
coincident with the brightest X-ray sources (see Fig.1 in Antokhin \e
2008), and are those exerting the strongest influence on their
neighborhood. Accordingly, we define five spatial regions, delimited by
red dashed segments in the Figure, centered on stars HD93129A
(O2If*$+$O3.5, central star of Tr~14), V560 Car (= HD93205,
O3.5V((f$+$))$+$O8V), HD93250 (O4IIIfc:, central star of Collinder~232),
WR25 (= HD93162, O2.5If*/WN6), and $\eta$ Car.

\section{Reddening}
\label{reddening}

Determination of optical extinction towards individual stars is very
important to establish how they are distributed along the line of sight.
A number of studies (e.g., Smith 1987, Hur \e 2012) have
determined that the foreground extinction towards Carina is relatively low
($E(B-V) \sim 0.36 \pm 0.04$, or $A_V^{fg} \sim 1.1$),
while higher-reddening stars in the same region
show an anomalous reddening law with $R=4.4-4.8$.
In the central part of Carina being studied here, highly obscured member stars
have been found by means of X-ray observations (Albacete-Colombo \e
2008; CCCP).
Therefore, even considering only
Tr~14/16 (and not the embedded clusters of
Feigelson \e 2011), cluster members are found over a large range of
extinction values. The spatially highly non-uniform distribution of
extinction is also made obvious by the two prominent SE and SW dark dust lanes.

Figure~\ref{bv-vi-bstars} shows a $(B-V,V-I)$ color-color diagram
(using  photometry from Hur \e 2012), where in particular the colors of the
early-type stars (large circles) found from our spectroscopy are useful
to study the intra-cluster reddening law. The intrinsic colors of
massive stars (down to A0) are described by the red line (3-Myr
non-rotating solar-metallicity isochrone from Ekstr\"om \e 2012,
henceforth Geneva\footnote{
http://obswww.unige.ch/Recherche/evoldb/index/Isochrone/}), while those of
lower-mass stars by the grey line (ZAMS from Kenyon and Hartmann 1995,
henceforth KH95).  Despite the lack of a detailed
classification for the massive stars (Sect.~\ref{ob}), it is expected
that most of them will be late-B/early-A stars, with colors $B-V \sim
V-I \sim 0$. On this basis, the slope of the dashed arrow in
Fig.~\ref{bv-vi-bstars} suggests that the foreground reddening law is
$E(V-I)/E(B-V) = 1.53$ (slightly larger than the value 1.32 in Bessell \e 1998),
while inside Carina this ratio becomes 1.95 (slope of the black solid
arrow), slightly larger than the
value found by Hur \e (2012) of $1.8 \pm 0.1$ from photometry alone.
Figure~\ref{bv-vi-bstars} also shows that some of the B-type stars
redder than $B-V \sim 1$ appear to have a lower $E(V-I)/E(B-V)$ ratio,
more like a normal reddening law, although their spatial distribution
shows no obvious pattern. Moreover, the two candidate O stars found in
Sect.~\ref{ob} are both found at very large reddenings (orange and
violet circles), corresponding to intra-cluster extinction $A_V>5.5$.
Among stars with $B-V \geq 2$, the mentioned O star is the only
candidate cluster member: the handful of other stars found in that color
range are therefore obscured background giants, whose detection permits
us to estimate the total optical
{ extinction}
of the Nebula, and its spatial variation, as we will discuss below
{ and in Section~\ref{structure}}.

We have determined individual extinction values for our stars from their
photometric $V-I$ colors, \teff\ derived from our spectroscopy, and 
the \teff-color relation from KH95 for ZAMS stars\footnote{Although the
KH95 relation may be superseded by more recent calibrations, its usage
here for deriving $A_V$ is appropriate for best consistency with the
\teff\ calibration in Damiani \e (2014).};
the $E(V-I)$ colors
were converted to $A_V$ using the above reddening laws as appropriate, i.e.\
differently for foreground extinction only ($A_V<1.1$) and
foreground$+$intra-cluster extinction, each with its reddening law.
Because of our inability to assign a detailed spectral type and
photometric color to the early-type stars, their extinction was computed
by de-reddening their photometry to match the 3-Myr isochrone from
Ekstr\"om \e (2012).
Figure~\ref{teff-vi-groups} shows, separately for the different spatial
groups (defined in Section~\ref{spatial}), \teff\ vs.\ $V-I$
diagrams for all of our stars. The solid line is the KH95 calibration at
zero reddening, while the dashed line is the same curve at $A_V =
A_V^{fg} = 1.1$; it is clear that most low-mass cluster members fall
a little to the right of the latter curve. In this Figure we also show the
placement of giant stars (defined from the $\gamma$ index in Damiani \e 2014)
as red circles: as anticipated above, most of
these giants lie at large reddening values\footnote{The employed KH95
calibration is appropriate to ZAMS stars, and therefore the reddening
values derived by applying it to giants are only approximate, but
sufficient for our classification as background stars.},
and likely beyond the Carina Nebula.
{
In the Trumpler~14 region virtually no background stars are seen: this
might be related to a relatively higher extinction compared to other
sightlines, and also to our incomplete target
sampling combined with a locally enhanced member-to-field star density ratio.
}

\section{Color-magnitude diagram}
\label{cmd}

\begin{figure}
\resizebox{\hsize}{!}{
\includegraphics[bb=5 10 485 475]{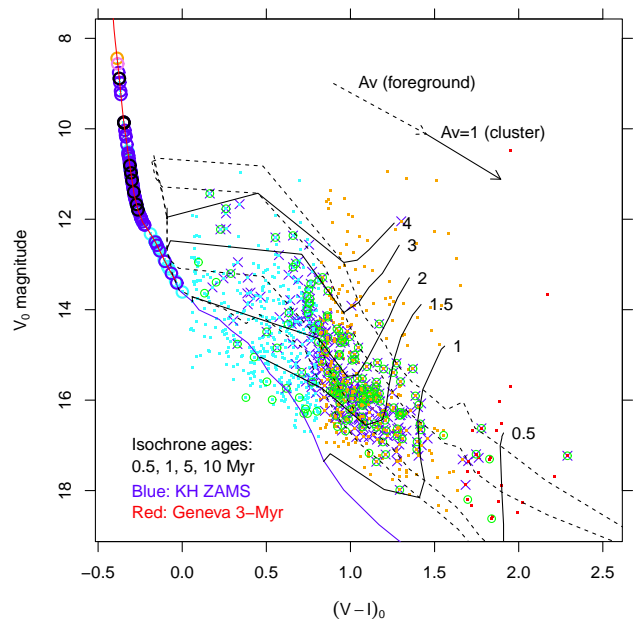}}
\caption{Dereddened $(V,V-I)$ color-magnitude diagram.
Symbols as in Figs.~\ref{teff-li} and~\ref{ra-dec-bstars}.
Also shown are the ZAMS for lower MS stars (blue), a Geneva 3-Myr isochrone for
massive stars (red), and selected isochrones (dashed black) and evolutionary
tracks (solid black) for lower-mass stars from Siess \e (2000);
these latter are labeled with
stellar mass in solar units. A0 and earlier-type stars have been
projected to the MS
{ (along the combined foreground$+$cluster reddening vector, as shown),
since they lack a detailed classification.}
\label{v-vi-0}}
\end{figure}

We show in Figure~\ref{v-vi-0} the dereddened CMD of our sample stars,
We find probable cluster members of
masses down to one solar mass (or slightly below).
The nonuniform reddening however prevents
from establish a mass completeness limit: in places where reddening is
largest we cannot be sure of detecting even O stars (our candidate O
star [S87b] IRS 38 has $V=17.64$). Nonuniform reddening causes
bright stars to be over-represented among optically selected samples
such as this one, hence the relatively large number of B stars relative to
solar-mass stars we find in our sample. This dataset is therefore
unsuitable for studies (e.g.\ of the IMF) requiring statistically
representative samples.
From the Figure it is also apparent the scarcity of cluster members in
the $(V-I)_0$ range 0.2-0.5, corresponding to late-A and F stars. The
same lack of members can be observed in Fig.~\ref{teff-vi-groups} at
\teff\ near 7000~K. Since lower-mass stars are found in larger numbers,
this cannot be due to excessive reddening towards the F stars. One
possible explanation for the small number of F-type members is the
relatively fast traversal time across their radiative tracks in the CMD.
In addition, we might lack the ability to assess their membership, 
as mentioned already in Sect.~\ref{membership} for the A-type stars. Since
F-type stars are generally found to be as strong X-ray emitters as
G-type stars, we will consider again them in the context of the Carina
X-ray data below.

Ages for the low-mass stars estimated from the isochrones in
Fig.~\ref{v-vi-0} are
distributed mostly in the range 1-5 Myr, in agreement with previous
studies (e.g., Tapia \e 2003, Carraro \e 2004, Hur \e 2012). 
As noted by e.g.\ Smith and Brooks (2008) or Wolk \e (2011), not all
clusters in Carina have the same age (Tr~14, in particular, is more
compact and probably younger than Tr~16).
The problem of establishing their age relationship will be discussed in
Section~\ref{ages}.

\begin{figure}
\resizebox{\hsize}{!}{
\includegraphics[bb=5 10 485 475]{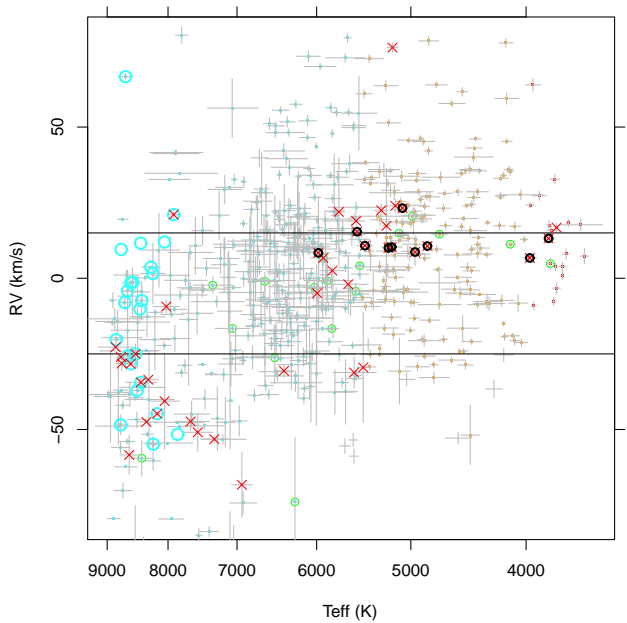}}
\caption{RV vs.\ \teff\ for stars not classified as members.
Symbols as in Figure~\ref{teff-li}, with the addition of red crosses
representing X-ray detected stars, unconfirmed by other membership indicators.
Bars indicate statistical errors only.
The RV10 stars are indicated with black circles.
The horizontal lines indicate the cluster RV range.
\label{teff-rv-nonmembers}}
\end{figure}

\section{X-ray data}
\label{xray}

In the subregion comprising Tr~16 and WR25 we find approximately 180 low-mass
candidate Carina members.
{ Considering that our target selection
involved a $\sim 50$\% down-sampling of all candidate members from
optical photometry down to $V=18.5$, we can extrapolate to $\sim 360$ members
down to that magnitude limit: this is much smaller than the number of X-ray
sources (1035) found in the same region by Albacete-Colombo \e (2008).}
An even
larger number was found in the CCCP X-ray survey, although based on the
same Chandra observation in the field of Tr~16, because of the less
conservative choices adopted in detecting point sources. Spurious
detections in Albacete-Colombo \e (2008) are predicted not to exceed
$\sim 10$ sources, and the number of unrelated field X-ray sources is
$\sim 100$,
{ so that the number of X-ray members in this subregion remains
$>900$.}
Therefore, a significant excess of $\geq 500$ X-ray sources
remain, above the number of members inferred from our spectroscopic
data. These must be stars fainter than our $V=18.5$ selection limit,
either because of their low mass, or because of large extinction (or
both). Since their number is not small with respect to the
spectroscopically studied sample, we cannot overlook them for a proper
understanding of the properties of the SFR.
{ Here we try to investigate their nature.}

{ Fourty X-ray detected late-type stars (21 from Albacete-Colombo \e 2008)}
were observed spectroscopically, but not
classified here as members;
{ of them, 11 are RV10 stars, and 6 are near-ZAMS rejected members
discussed in Section~\ref{membership}.}
We first examine whether this can reveal a failure of our membership criteria.
These stars do not cluster at
any particular place in the CMD diagram, and there is no reason to conclude
that as a group they should be included among members. The same
conclusion is reached from Figure~\ref{teff-rv-nonmembers}, showing RV
vs.\ \teff\ exclusively for non-member stars: the X-ray emitting
non-members are randomly scattered, and in particular their membership
status would not change by any small adjustment of the nominal RV range
for members (horizontal lines). We recall that, if these X-ray sources
had lithium EW above 150~mA, they would have been accepted as members
even in the presence of discrepant RV (in order not to lose
binaries)\footnote{
The few red crosses inside the nominal RV range were rejected as members
because of their proximity to the ZAMS, as explained above.}.
Thus, the red crosses in Fig.~\ref{teff-rv-nonmembers} have not only
discrepant RV but
also no lithium. As discussed above, in the \teff\ range 9000-7000~K
our membership criteria may be weakest; yet, even in this range
the number of potential candidates from X-rays is very small, and they
appear to have all significantly discrepant RV.
Therefore, these X-ray detected stars with no lithium and discrepant RV
are unlikely members, and will not be considered further\footnote{
{ Of these 40 stars, 37 are classified by Broos \e (2011b) within the
CCCP project, on the basis of X-ray properties and optical/NIR photometry
alone: all but four (unclassified) objects were assigned to the ``Carina
young star'' group, including thus the majority of near-ZAMS and RV10
stars, which we instead excluded on the basis of the Hur \e (2012) photometry,
and our RVs, respectively.}}.
We also conclude that the lack of member stars in the temperature range
9000-7000~K is real, and not a byproduct of our membership criteria.
{
This is probably related to the R-C gap found in the photometric CMD
of several clusters by Mayne \e (2007); the same effect is not
recognizable in the CMD of Carina clusters because of strong differential
extinction.
}

\begin{figure*}
\includegraphics[bb=5 10 485 475,width=9cm]{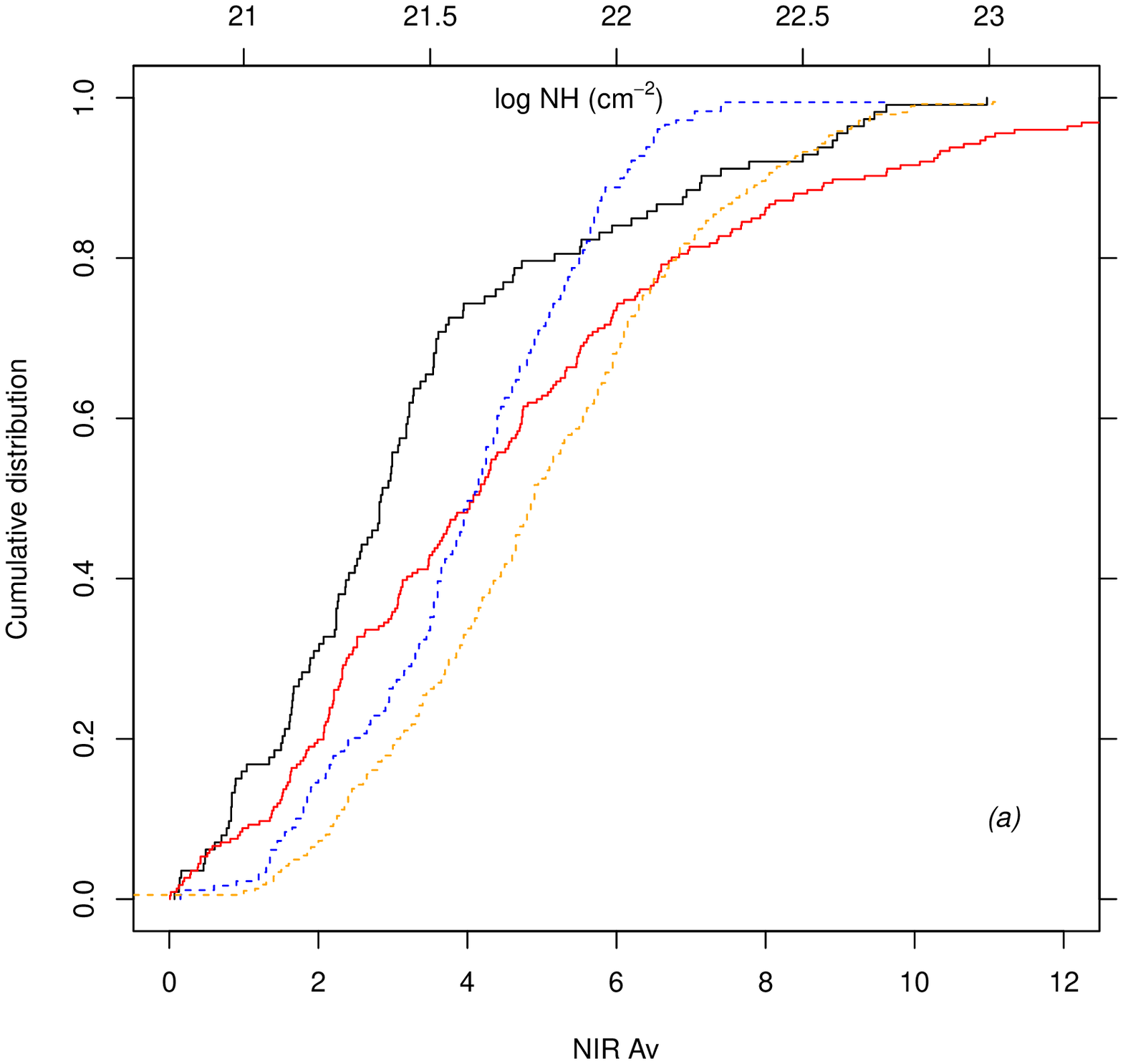}
\hfill
\hfill
\includegraphics[bb=5 10 485 475,width=9cm]{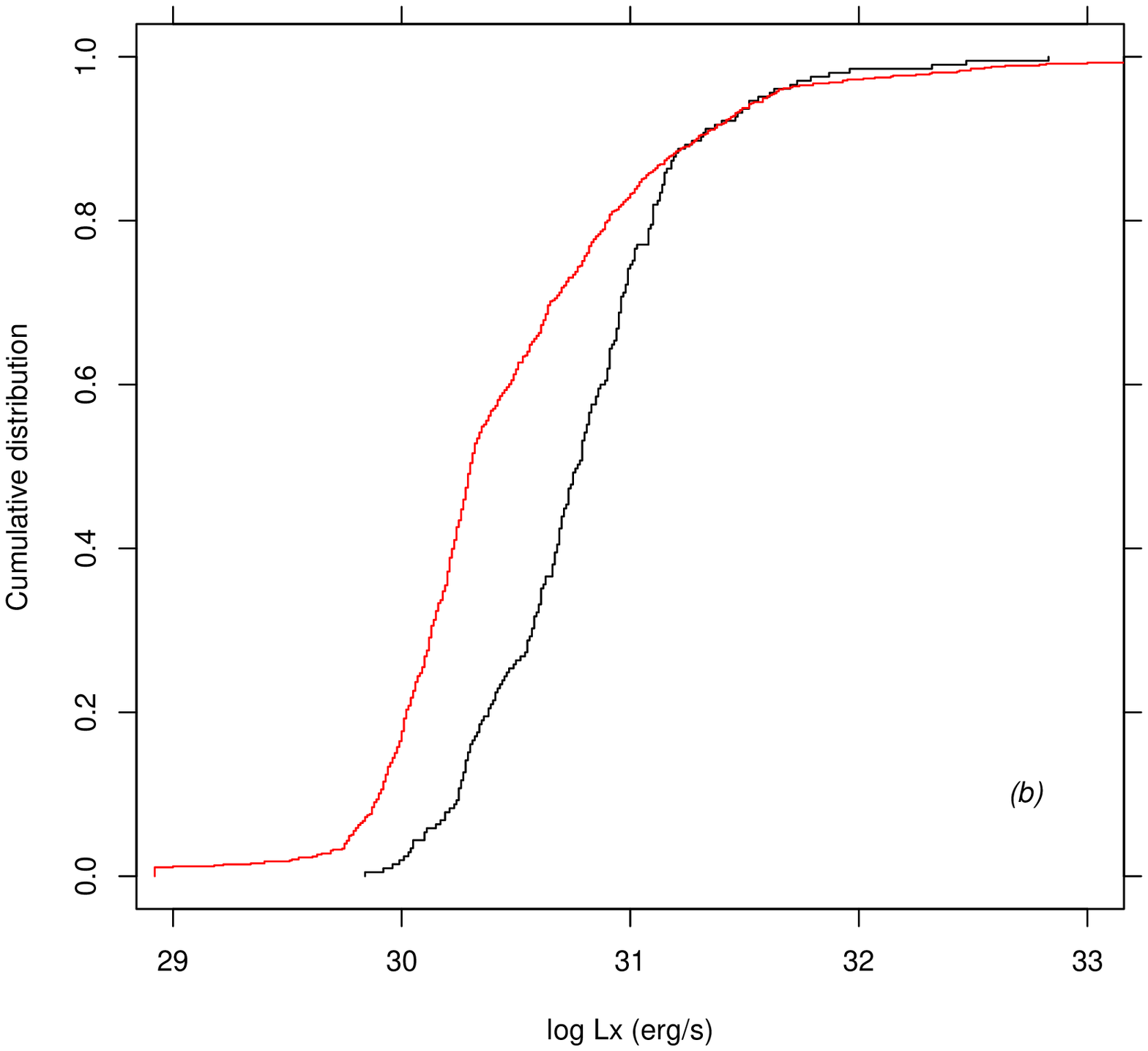}
\caption{
$(a)$:
Solid: Cumulative distributions of $A_V$ (derived from NIR colors), for
Tr~16 X-ray sources observed here spectroscopically (black) and unobserved
(red).
Dashed: Cumulative distributions of gas column density $N_H$ (scale on
top axis), for the same subsamples (blue and orange respectively).
$(b)$:
As in panel $(a)$, for the X-ray luminosity $L_X$.
\label{cumdist-xsrc}}
\end{figure*}

Next, we examine the X-ray properties of X-ray detected sources which
have not been observed spectroscopically. We take the list of sources
and their properties from Albacete-Colombo \e (2008).
Figure~\ref{cumdist-xsrc}-$a$ compares cumulative distributions of
extinction $A_V$ (as derived from the near-IR colors, for X-ray sources
with a 2MASS counterpart), for the X-ray sources in the spectroscopic
sample and in the no-spectroscopy sample. A Kolmogorov-Smirnov (KS)
statistical test gives a
probability $P$ that the two distributions are
{ drawn from the same parent sample}
of only $P= 4.9 \cdot 10^{-5}$, confirming the
significance of their visually apparent difference.
Similar results are obtained from consideration of the absorbing column
density $N_H$, as derived from model best fits to the Chandra ACIS X-ray
spectra (Albacete-Colombo \e 2008), whose distributions for the
spectroscopically observed and unobserved
samples of X-ray sources are also shown in Figure~\ref{cumdist-xsrc}-$a$:
here a KS test gives a probability of $P= 1.83 \cdot 10^{-6}$ that the two
distributions are { drawn from the same parent sample}.
The two tests just made do not refer exactly to the same subsamples of
X-ray sources, since only 2/3 of the X-ray sources in Albacete-Colombo \e
(2008) have a 2MASS counterpart, and only those detected with more than
20 X-ray counts had a spectral fit performed.
{ Nevertheless,} there is definite evidence that the optically missed, X-ray
detected members are more obscured than the stars in our spectroscopic
sample.

\begin{figure}
\resizebox{\hsize}{!}{
\includegraphics[bb=5 10 485 475]{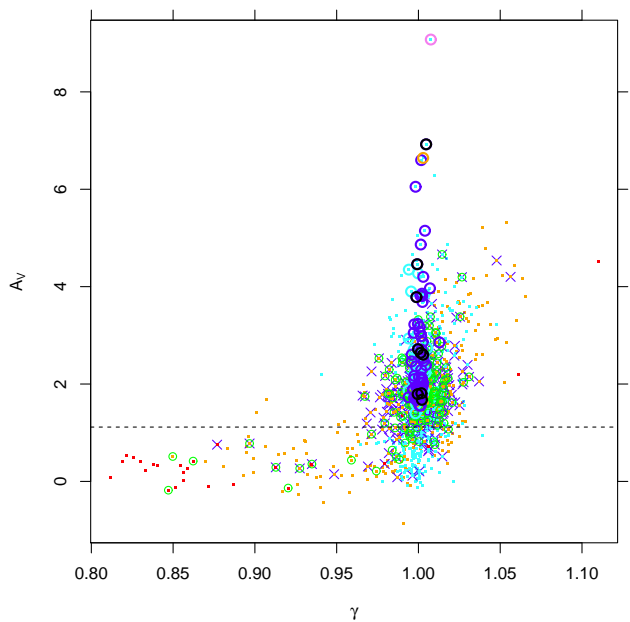}}
\caption{Optical extinction $A_V$ vs.\ spectral index $\gamma$.
Symbols as in Figs.~\ref{teff-li} and~\ref{ra-dec-bstars}.
For GKM stars, $\gamma$ is an effective gravity diagnostics, with
high-gravity stars having $\gamma<0.97$ and giants having $\gamma>1.02$.
The horizontal dashed line indicates the foreground extinction value.
\label{gamma-av}}
\end{figure}

\begin{figure*}
\centering
\includegraphics[bb=0 0 576 576,width=17cm]{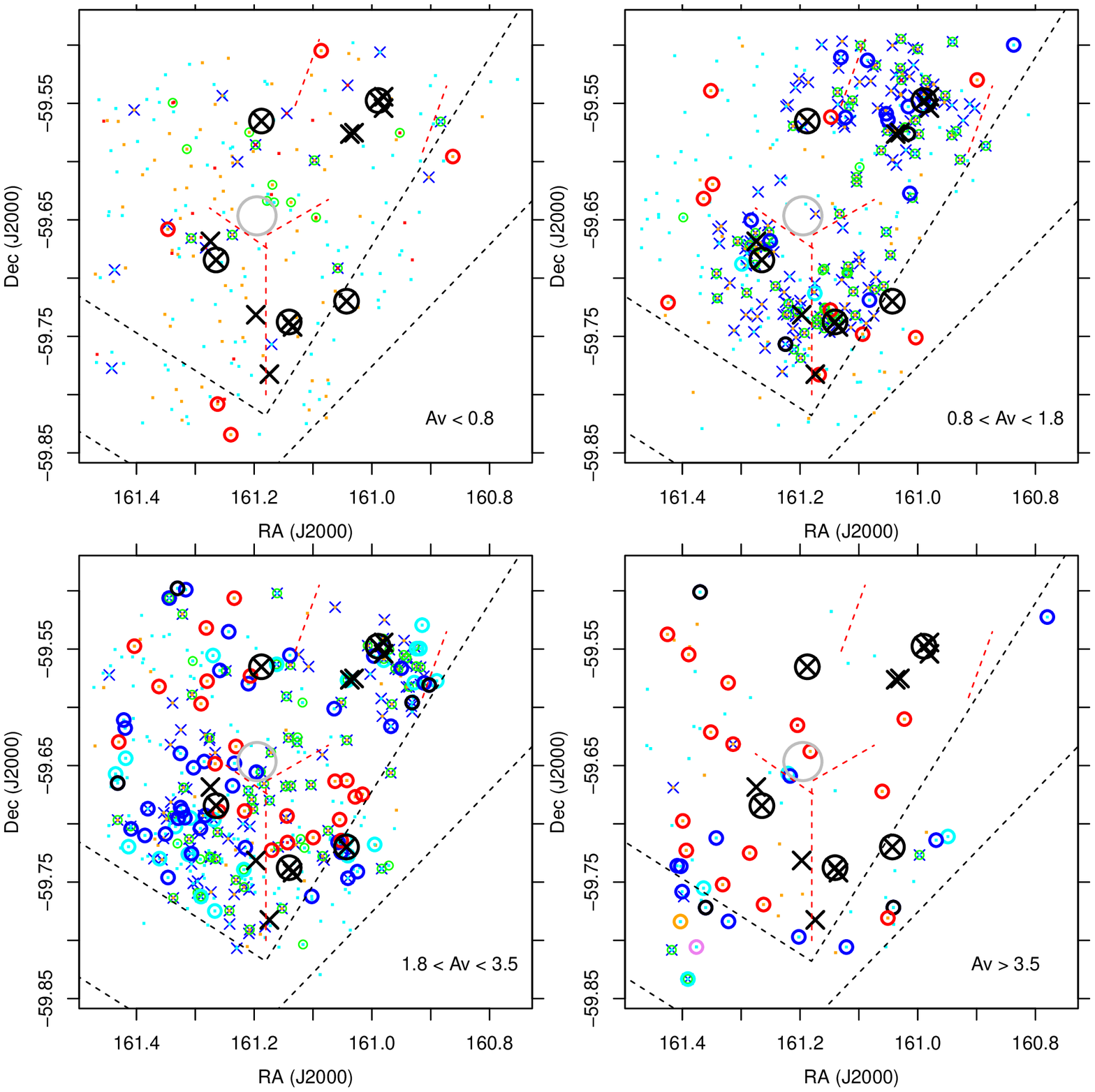}
\caption{Spatial maps of observed stars, split in four ranges of $A_V$
as labeled. Star symbols as in Fig.~\ref{ra-dec-bstars}, with the
addition of red circles indicating low-gravity giants as in
Fig.~\ref{teff-vi-groups}. Dashed black lines indicate the dark dust
lanes as in Fig.~\ref{ra-dec-bstars}. Dashed red segments indicate the
boundaries of large gas shells studied in Paper~I. The big grey circle
indicate position and approximate size of the Keyhole nebula.
\label{ra-dec-av}}
\end{figure*}

This finding does not rule out that the X-ray detected sample may
contain also a number of low-extinction members at fainter optical
magnitudes than our limit. To test this, we consider the distributions
of X-ray luminosities $L_X$, derived from X-ray spectral fits and
therefore corrected for absorption, for the spectroscopic and
no-spectroscopy samples, as shown in Figure~\ref{cumdist-xsrc}-$b$:
the KS test gives a probability of no difference of $P= 2.2 \cdot 10^{-16}$.
Young PMS low-mass stars are known to be in an X-ray saturated regime,
where the X-ray luminosity $L_X$ is on average proportional to stellar
bolometric luminosity, $L_X \sim 10^{-3} L_{bol}$ (Flaccomio \e 2003,
Preibisch \e 2005, Damiani \e 2006a); therefore, the lower (unabsorbed)
X-ray luminosities of stars without spectroscopy implies that these
stars have on average lower luminosities (and mass) than our
spectroscopic sample. The X-ray detected member sample is therefore more
complete than the spectroscopic sample both towards lower masses and
towards more obscured stars.

\section{Cluster structure}
\label{structure}

We attempt here to combine all data discussed above in a single coherent
picture of the structure of central Carina. We first test the
meaningfulness of the derived extinction values $A_V$, by comparing them
with a spectroscopic gravity index $\gamma$, defined in Damiani \e
(2014). The prediction being tested here is that all main-sequence
stars later than mid-G, being only observable in the foreground of the
Nebula, must have $A_V \leq A_V^{fg}$, while most giants will appear in
the background, having a much lower space density but higher luminosity.
Figure~\ref{gamma-av} shows that this is exactly the case: 
high-gravity, main-sequence GKM stars have $\gamma<0.97$, and the vast
majority of them is found at $A_V \leq A_V^{fg}$, while giants
(having $\gamma>1.02$) are almost all found in the $A_V$ range between
2-5 mag, typically larger than the extinction towards cluster members.

\begin{figure}
\resizebox{\hsize}{!}{
\includegraphics[bb=5 10 485 475]{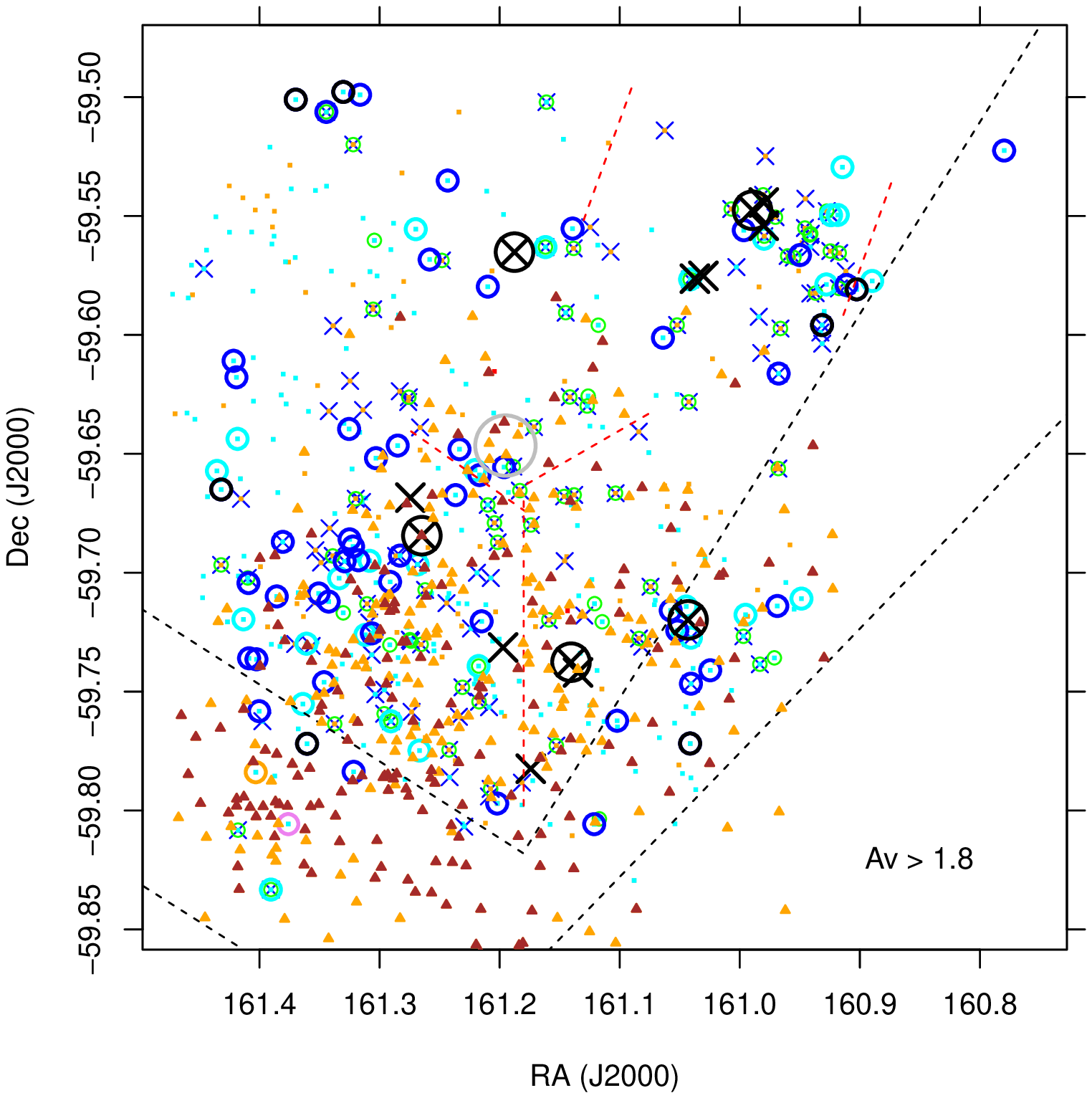}}
\caption{Spatial map of reddened sample stars with $A_V>1.8$ (shown by
same symbols as in Fig.~\ref{ra-dec-av}), and of obscured X-ray sources
(orange triangles: $0.5<HS<1.0$; brown triangles: $HS>1$).
\label{ra-dec-xray-obsc}}
\end{figure}

Having better established our confidence in the determined extinction
values, we next consider the spatial distributions of stars in several
ranges of $A_V$, as shown in Figure~\ref{ra-dec-av}.
The upper left panel shows all stars with $A_V<0.8$:
no clustering
is detectable, in agreement with these being foreground stars, unrelated
to Carina and its obscuring material. Tens of stars are found projected
against the dark V-shaped dust lanes. The number of low-gravity stars
(red circles) is here very low. In the upper right panel, the
low-extinction cluster members appear, in dense groups with close spatial
association to the most massive stars; the dust
lanes are here almost devoid of stars, except for a dozen (probably background)
stars near WR25, where the total absorption is lower than elsewhere in
the lane (see e.g.\ Fig.1 in Albacete-Colombo \e 2008). Most of the
B-type stars have not yet made their appearance in this $A_V$ range:
this makes the number ratio between the early-type and the solar-type
stars here probably more representative of its real value.
In the next extinction range, shown in the lower left panel, the stellar
distribution has changed considerably: low-mass members are more widely
dispersed away from the most massive stars, while their density generally
increases towards places of higher obscuration. This is seen in the
vicinity of $\eta$~Car, which has now many more neighbors towards SE
(where they eventually meet the dark lane) than towards NW, and also in
the Trumpler~14 region, where again member stars are found exclusively
on the side of the cluster nearest to the dark lane. Taking the dashed
black lines in the Figure as a reference, we may observe that the number
of stars in their immediate vicinity, in both the Tr~14 and $\eta$~Car
subregions, increases dramatically from the low-extinction layer to the
higher-extinction one. This is also true for the B/early-A stars, most
of which found on the high-extinction sides of Tr~14 and $\eta$~Car,
respectively. The South-West and North-East regions, around WR25 and
HD93250/Collinder~232 respectively, show several low-gravity background
giants, indicating that the total Nebula extinction in those regions
does not exceed $A_V \sim 2-3$. No background giants are instead found
on the high-extinction sides of Tr~14 and $\eta$~Car, nor in the dark
lanes (except for the vicinity of WR25 as already remarked).
Finally, the lower right panel shows the most reddened stars found in
our dataset: only early-type cluster members are found here, including
the two candidate O stars (both under the SE obscuring lane) and three
candidate B supergiants; at these extinction values, background giants
make their appearance even close to the SE dark lane,
{ but not in the region surrounding Tr~14.}
This extinction pattern agrees qualitatively with that in Smith
and Brook (2007; their Fig.5$a$).

Figure~\ref{ra-dec-xray-obsc} shows all stars with $A_V>1.8$ together
with X-ray detected sources in Tr~16 (most of which members as discussed above)
with X-ray hardness index $HS>0.5$\footnote{$HS$ is defined as the count
ratio between bands [2-8]~keV and [0.5-2]~keV.}, and therefore moderate
to large absorption,
and not observed spectroscopically. Most moderately-obscured X-ray sources
(orange triangles) lie South of $\eta$~Car and between it and WR25,
while the heavily-obscured X-ray sources (brown triangles) lie behind
the sourthern dark lane, with the largest source density found between
the two new candidate O stars. This subcluster of highly absorbed X-ray
sources is named CCCP-Cl~14 in Feigelson \e (2011), and Tr~16 SE in
Sanchawala \e (2007). Its dominant star, according to Feigelson \e (2011),
is the O eclipsing binary FO~15
(= V662~Car, O5.5Vz + O9.5 V,v), which however we did not observe
spectroscopically. The optical extinction of this star is $A_V \sim 5$
(Niemela \e 2006), thus lower than that we find for our
candidate O stars [S87b] IRS 42 and IRS 38 ($A_V = 6.6$ and 9.1,
respectively) in the same region.
At the same location, an obscured cluster of massive stars was found
from Spitzer images by Smith \e (2010 - called 'cluster G' in their
Table~4), who argued that this is not an embedded, extremely young cluster,
but more likely a cluster of age 1-2~Myr, behind and not inside the
foreground dark lane.

Figure~\ref{ra-dec-xray-obsc} shows no
clear separation between subclusters, in proceeding from $\eta$~Car to
its SE, but only a gradually increasing proportion of highly absorbed
Carina members, both massive and of low-mass (the latter only from
X-rays). Instead, the low-extinction members (in Figure~\ref{ra-dec-av},
upper right panel) are much more clustered in tight groups centered on the
respective massive stars.
It is interesting to compare the distribution of stars with that of the
ionized gas from Paper~I, consisting of several large shells, almost
devoid of dust, centered on $\eta$~Car, WR25, and Tr~14.
The observed gradual increase in extinction towards SE, if most stars are
enclosed in such dust-free shells, implies that the obscuring material
must lie in front of the star clusters (and gas), and not generally mixed with
them.
At its SE border, the
$\eta$~Car gas shell does not seem to interact with matter in the dark lane,
but instead fades behind it: this agrees with the
conclusion by Smith \e (2010) that the obscured SE subcluster (their 'cluster
G') lies behind the lane and does not interact with it.
This is probably true also of the newly found massive OB stars in the
direction of the dark lanes, so that their UV radiation would illuminate
most cluster stars with little attenuation, despite the very large
extinction found towards us.
Of course, circumstellar dust may still exists around individual stars,
which contributes to the $A_V$ scatter in Fig.~\ref{ra-dec-av} and
shields the massive-star UV flux; however, since we do observe a regular
spatial extinction pattern in Fig.~\ref{ra-dec-av}, local extinction is
probably a minor contributor to the total observed $A_V$.

Behind the high-reddening members of Tr~14, with $A_V \leq 3$,
{ we do not find virtually any background star.}
Except for the Tr~14 region and the dark lanes, we have instead found
background giants throughout most of the surveyed part of Carina,
and therefore we have been observing the brightest members through the entire
thickness of the nebula until its far side.

\section{Stellar ages}
\label{ages}

We investigate here if there are measurable age
differences between the Carina subclusters.
An useful age diagnostics for young low-mass stars is
the lithium resonance
line at 6707\AA, whose EW as a function of \teff\ was shown in
Fig.~\ref{teff-li}.
In the temperature range of the Carina members studied here, however,
lithium EW is insensitive to age if a cluster is younger than several tens
Myr (Jeffries 2014), as is the case for Carina clusters.  Accordingly,
the scatter of datapoints in the above figure may be entirely accounted
for by errors on \teff\ ($\sim 5$\%) and lithium EW, and is not related
to real age differences among the observed Carina members.

\begin{figure}
\resizebox{\hsize}{!}{
\includegraphics[bb=5 10 485 475]{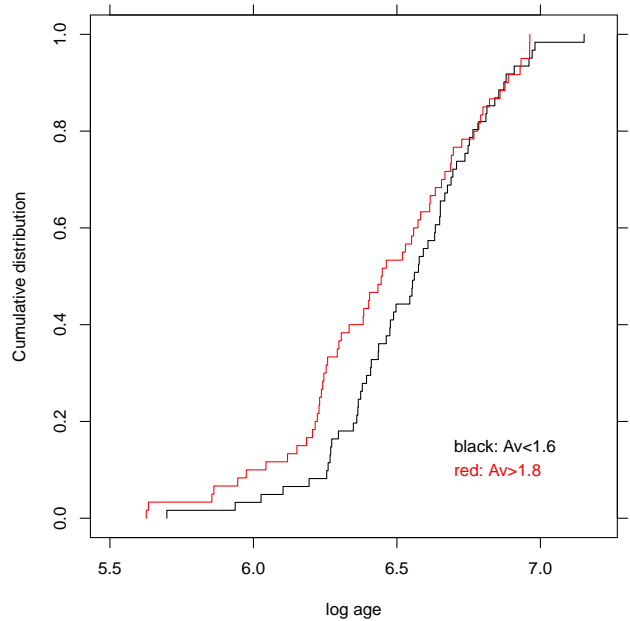}}
\caption{Cumulative distributions of photometric ages, for all members
in the \teff\ range 4300-6500~K.
Black lines: $A_V<1.6$; red lines: $A_V>1.8$.
\label{age-cumfun-av}}
\end{figure}

\begin{figure}
\resizebox{\hsize}{!}{
\includegraphics[bb=5 10 485 475]{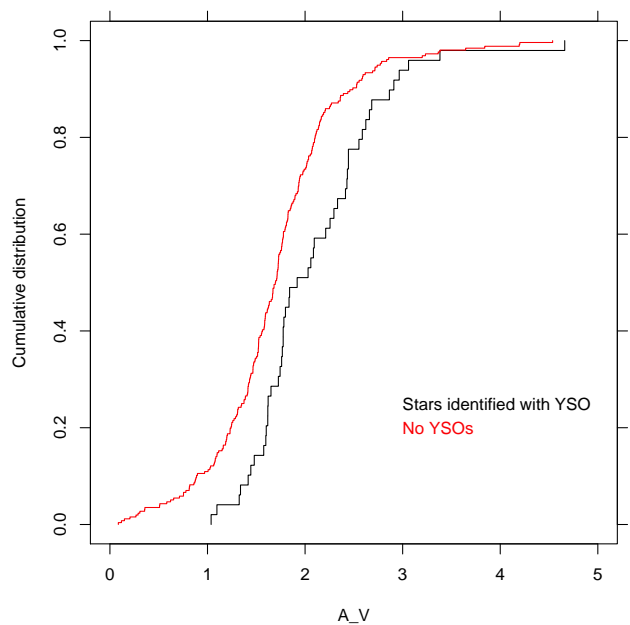}}
\caption{Cumulative $A_V$ distribution for members identified with a
YSO from Zeidler \e (2016) (black line), and for those without a YSO
counterpart (red).
\label{cumdist-av-yso}}
\end{figure}

\begin{figure}
\resizebox{\hsize}{!}{
\includegraphics[bb=5 10 485 475]{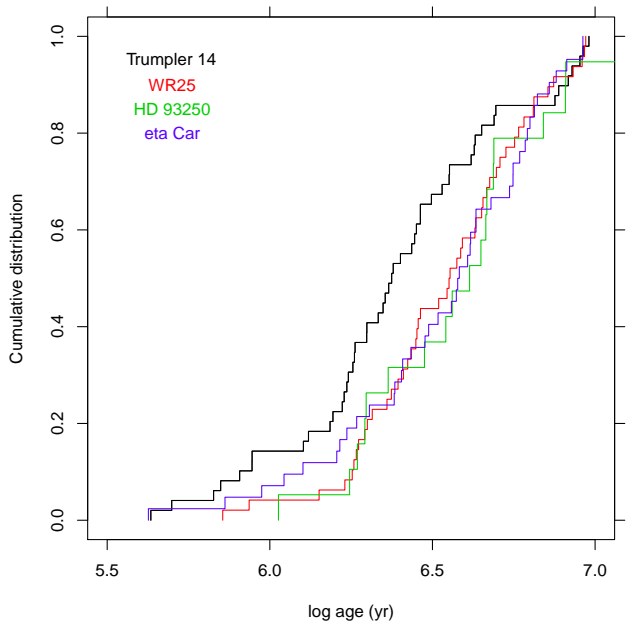}}
\caption{Cumulative distributions of stellar ages as derived from
photometry and isochrones, as in Fig.~\ref{v-vi-0}, for cluster
members from different subgroups.
\label{age-cumfun-groups}}
\end{figure}

We examine the relative ages of the different Carina
subgroups as derived from the star location in the CMD, compared to
Siess \e (2000) isochrones. This methods still benefits from our
spectroscopic data since apparent star colors and magnitudes are
individually de-reddened using the extinction values $A_V$ derived
above. Besides that, there remain a series of caveats related to
non-photospheric contributions such as veiling in accreting PMS members,
uncertainties in the reddening law, or unrecognized binaries.

The distributions of photometric ages for low- and high-extinction
subsamples are shown in Figure~\ref{age-cumfun-av}: both distributions
are rather wide, over more than one dex in ages.
This agrees with the earlier result of DeGioia-Eastwood \e (2001) that
star formation in these clusters was active over the last 10~Myr, which
should be taken as an upper limit to the true age spread considering
e.g.\ uncertainties in extinction and reddening law, binarity, or variability.
The difference between the two distributions in
the Figure points to the obscured Carina population being slightly
younger than the unobscured one, however with a modest statistical
significance of 93.9\%.

To corroborate this result, we show in
Figure~\ref{cumdist-av-yso} the cumulative distributions of $A_V$ for
Carina members with and without a YSO counterpart in the Zeidler
\e (2016) catalog: (younger) members associated with a YSO
have a larger extinction than (older) members with no YSO association,
with a confidence level of 99.6\%.

The cumulative distributions of photometric
ages of Carina member stars in the subclusters are shown in
Figure~\ref{age-cumfun-groups}: stars in Tr~14 are significantly younger
(at 99.1\% level) than those of all other subgroup (cumulatively), while the age
distributions of $\eta$~Car, HD93250 and WR25 subgroups are
indistinguishable from one another. This agrees with the already mentioned
suggestion that Tr~14 is younger than Tr~16.

\section{Rotation}
\label{rotation}

\begin{figure}
\resizebox{\hsize}{!}{
\includegraphics[bb=5 10 485 475]{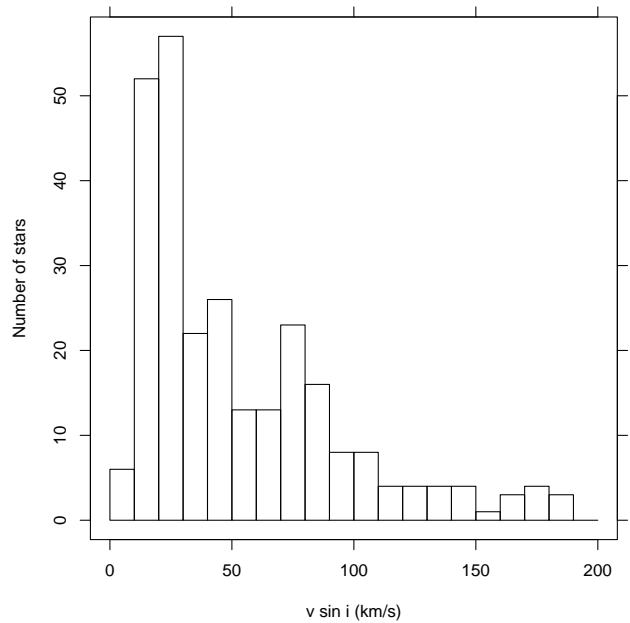}}
\caption{$v \sin i$ distribution for Carina members with
$4500 <T_{eff} <6500$~K.
\label{vrot-hist}}
\end{figure}

\begin{figure}
\resizebox{\hsize}{!}{
\includegraphics[bb=5 10 485 475]{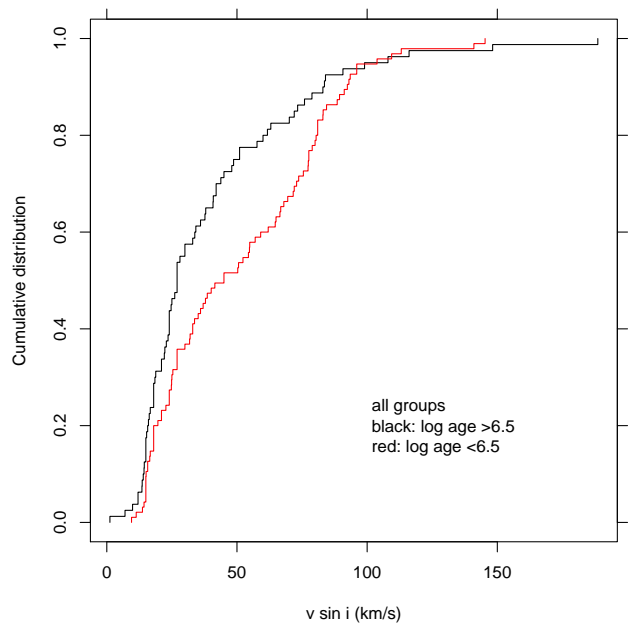}}
\caption{Cumulative $v \sin i$ distributions for low-mass member stars from all
groups, split by photometric age. Black: log age $>6.5$;
red: log age $<6.5$ (yr).
\label{cumdist-vrot-age}}
\end{figure}

The HR15n spectra of Carina members show that a
large fraction of these stars, including lower-mass ones, are fast
rotators, with a median $v \sin i \sim 40$~km/s for $4500< T_{eff} < 6500$~K
(Figure~\ref{vrot-hist}).
To our knowledge,
these are the first measurements of rotation rates for Carina PMS stars.
Since average stellar rotation rates are strongly mass-dependent, we have
chosen for the above Figure a restricted \teff\ range.
Moreover, the Figure suggests a bimodal distribution, with a higher peak
near $v \sin i \sim 22$~km/s, and a secondary
one at $v \sin i \sim 75$~km/s, qualitatively similar to that in the
ONC (Stassun \e 1999, Herbst \e 2002) for masses larger than $0.25 M_{\odot}$.
Assuming a typical radius for our Carina PMS stars of $2 R_{\odot}$, the
$v \sin i$ peaks would transform into rotational-period peaks around 1.1
and 3.6~days, respectively, not significantly different from those found in
the ONC.
We do not find a significant difference between the rotational velocity
distributions of CTTS and WTTS, neither globally nor considering each
subgroup separately; however, as we have discussed in
section~\ref{membership}, an accurate assessment of CTTS status is
difficult using our spectra. We have therefore examined the $v \sin i$
distributions for stars with and without NIR excesses, as measured using
the 2MASS $(H-K,J-H)$ color-color diagram (as e.g.\ in Damiani \e 2006b),
or the Spitzer IRAC $([3.6-4.5],[5.8-8.0])$ diagram (as in Povich \e 2011b).
In our Carina sample only very few stars are found
with significant excesses in these NIR/MIR colors, and their
rotational-velocity distribution is not significantly different from
that of stars with no excesses. Last we have considered our Carina members
with a counterpart in the YSO catalogue of Zeidler \e (2016):
the $v \sin i$ distributions for stars associated and unassociated with a YSO
are different at the 92.5\% significance level.
This is marginal, probably because of the limited member sample size.
The sense of the difference, nevertheless, seems to indicate that stars
surrounded by a massive dust disk spin slower than stars with no disk,
giving some support to the disk-locking paradigm (e.g., Mathieu 2004,
Rebull \e 2006).

We find a more significant difference (at 98.55\% level) between
sample stars respectively younger and older than 3~Myr
(Figure~\ref{cumdist-vrot-age}).
The older stars rotate more slowly than younger ones.
This result is similar to that found by Littlefair \e (2011) for several
other young clusters.
This is
surprising for mostly diskless PMS stars, which should spin up as their
moment of inertia decreases during contraction. However, stars in this
Carina PMS sample belong mostly to the mass range $1-4 M_{\odot}$, and
nearly half of them lie along radiative tracks, where contraction is
much slower than along the Hayashi track (according to the Siess \e 2000
models, radius even increases along part of the radiative track for
some masses in this range). The richness of Carina, coupled to our
limiting magnitude, makes that the stellar mass composition of our
sample is very much complementary to that of most existing studies of
rotation in young clusters: we are unable to study the rotation of stars below
$\sim 1 M_{\odot}$, but have a sample of PMS stars in the $\sim 1-4
M_{\odot}$ range of a size hardly found in any other studied cluster.
Therefore, the Carina rotation data can be very important to study the
rotational evolution of PMS stars along their radiative track. Similar
datasets are still lacking for SFRs in the same richness class as Carina,
like those mentioned in Sect.~\ref{intro}.

Besides the difference between younger and older Carina members,
Fig.~\ref{cumdist-vrot-age} shows that the bimodal $v \sin i$
distribution is only found in the younger subsample: the older one does
not show two bumps in the cumulative distribution (the histogram peaks
correspond to the steepest derivatives in the cumulative distribution).
This may be related to
the development of the radiative stellar core, which has a profound influence
on the observed surface rotation (see e.g.\ Spada \e 2011).
Again, the Carina rotation data may be crucial for testing theoretical
models of rotational evolution along the radiative track.

\section{Discussion and conclusions}
\label{concl}

Our study of the Gaia-ESO dataset on Tr~14/16, the richest clusters in
the Carina complex, is to our knowledge the first extensive spectroscopic
exploration of a sizable sample of stars in a giant SFR, with a mass in
excess of $10^4 M_{\odot}$, of which Carina is a rare example.
Out of 1085 stars observed,
more than 100 turned out to be early-type stars, mostly lacking a
spectroscopic classification; among them are two new candidate O stars at
high extinction, already known as bright IR sources, 17 candidate
Herbig Ae/Be stars, and 9 candidate B supergiants, although the limited
wavelength range of the studied data prevented a more detailed
classification.
Based on RV, lithium, and X-ray data, we find 286 Carina low-mass members.
Their number ratio with respect to the massive stars is not
representative, because the wide range in extinction encountered favors
observation of bright member stars at high reddening.

We have confirmed the anomalous reddening law already
reported, and placed new constraints on the three-dimensional
space distributions of Carina members. In central Carina, there is a
relatively small percentage of embedded YSOs (Povich \e 2011); this and
other constraints posed by our previous study of the ionized gas
distribution (Paper~I) suggest a picture where the extinction towards
Carina young stars in the $\eta$~Car/Tr~16 region is mostly caused by dust at
some distance from the stars
themselves, not mixed with them. This has impact on the amount of UV
flux they receive from the most massive cluster stars, and on the
consequences that this is likely to have on the evolution of the
circumstellar disks (photoevaporation). It is worth noting that some
information on the local level of UV
irradiation may also come from a detailed analysis of some of the DIBs
found in the red spectra of these stars (Kos and Zwitter 2013).

The geometry of the studied region in Carina is not simple. We
observed background giants through several sightlines across the
studied region, behind a few magnitudes of optical extinction.
The dust lane to SE, close to $\eta$~Car, produces enough
obscuration to drive a candidate O star near to our limiting magnitude:
of course still more deeply obscured massive stars may exist in the same
direction. If a blister geometry applies to this part of the cloud,
this must be seen sideways from our sightline. In the Tr~14 region
close to the western dark lane, instead, the foreground obscuration is
a little more uniform, but rises { more sharply} towards the dust lane.
The two dust lanes have therefore a different
placement in space with respect to their nearest cluster.

The data suggest a complex history of star formation, with
a significantly younger age for Trumpler~14 with respect to all
other Carina subgroups.  The high-extinction stars are only slightly younger
than the low-extinction ones, and more frequently associated with a YSO.

Because of its richness, Carina also provides us with a sample of PMS
stars
in the $1-4 M_{\odot}$ range, of a size hardly found in other young
clusters.  We have presented $v \sin i$ distributions for this
unique
sample of stars (significantly larger than e.g.\ in the study of
intermediate-mass star rotation by Wolff \e 2004, in the ONC).
About one-half of these stars are evolving along radiative tracks.
We find evidence of a bimodal $v \sin i$
distribution, analogous to that found in the ONC for lower-mass stars.
Stars older than 3~Myr, and mostly on their radiative
PMS track, are found to rotate slower than younger stars, which puts
constraints on the rotational evolution of intermediate-mass PMS stars.

\begin{acknowledgements}
We wish to thank an anonymous referee for his/her helpful suggestions.
Based on data products from observations made with ESO Telescopes at the
La Silla Paranal Observatory under programme ID 188.B-3002. These data
products have been processed by the Cambridge Astronomy Survey Unit
(CASU) at the Institute of Astronomy, University of Cambridge, and by
the FLAMES/UVES reduction team at INAF/Osservatorio Astrofisico di
Arcetri. These data have been obtained from the Gaia-ESO Survey Data
Archive, prepared and hosted by the Wide Field Astronomy Unit, Institute
for Astronomy, University of Edinburgh, which is funded by the UK
Science and Technology Facilities Council.
This work was partly supported by the European Union FP7 programme
through ERC grant number 320360 and by the Leverhulme Trust through
grant RPG-2012-541. We acknowledge the support from INAF and Ministero
dell'Istruzione, dell'Universit\`a' e della Ricerca (MIUR) in the form
of the grant "Premiale VLT 2012". The results presented here benefit
from discussions held during the Gaia-ESO workshops and conferences
supported by the ESF (European Science Foundation) through the GREAT
Research Network Programme.
R.~B. acknowledges financial support from INAF under PRIN2013 Programme
'Disks, jets and the dawn of planets'.
This research has made use of the SIMBAD database,
operated at CDS, Strasbourg, France.
This work was also using data products from observations made with
ESO Telescopes at the
La Silla Paranal Observatory under programme ID 177.D-3023, as part of
the VST Photometric \ha\ Survey of the Southern Galactic Plane and Bulge
(VPHAS$+$, www.vphas.eu).
\end{acknowledgements}

\bibliographystyle{aa}

\begin{appendix}
\section{Sky-subtraction procedure}
\label{append}

{
The Gaia-ESO HR15n dataset on Tr~14/16 contains 185 pure-sky spectra from 137
distinct sky positions, taken simultaneously to the stellar spectra.
We have identified the following components in these sky spectra:
\begin{enumerate}
\item Sky-glow emission lines from Earth atmosphere: nearly identical
within spectra from the same Observing Block, variable with time (from
an Observing Block to the next), narrow within instrumental resolution,
at fixed wavelength in Earth frame.
\item Scattered solar light (typically moonlight): continuum with
absorption lines, being essentially a solar spectrum at very low
levels. Constant within the same Observing Block, but changing from one
Observing Block to the next.
\item Nebular \ha, He~I, [N II] and [S II] lines: they originate from
physically large regions near the target stars, and do not vary with time,
but have a strong spatial dependence in their strengths, line shapes,
widths, and velocities (see Paper~I).
\item Nebular continuum emission, from reflection nebulosity:
a non-negligible component in Tr~14/16, free of solar-like spectral
features but with a rather flat spectrum (like massive stars in the
nebula), time-constant but space-variable (Paper~I).
\end{enumerate}
The standard Gaia-ESO pipeline is able to remove components 1-2 (constant
within the same Observing Block, variable in time), but not to deal
with components 3-4.  Conversely, an attempt to correct a star spectrum
using only the sky spectrum nearest on the sky would have more success in
dealing with components 3-4, but would perform badly on components 1-2,
when these are significant.  Clearly, in the complex case of Tr~14/16
a combined approach is needed.

Stellar parameter evaluation relies on the depth of
stellar absorption lines/bands, so that an accurately determined stellar
continuum is of the greatest importance. Contamination by sky-glow or
nebular lines is in this context a minor problem, affecting very
localized wavelength regions/lines, which may be usually ignored in
deriving star parameters (except for the case of \ha).
We therefore focus first on sky continuum determination and correction.

In the Tr~14/16 dataset we find that the sky continuum does not
contain traces of scattered sunlight, using two methods. We computed
the cross-correlation function (CCF) of the sky spectra (nebular lines
excluded) with a solar spectrum, looking for any peak near RV=0, but no
such peak was found for any Observing Block. A second
method was to fit the sky spectrum with a constant, flat spectrum plus a
scaled solar spectrum: again, the scaling factor for the solar spectrum
was negligibly small with respect to the other, flat, component, for
all Observing Blocks.
Instead, within the same Observing Block the median sky continuum level
is found to vary significantly, and with a clear spatial pattern, a
clear indication of a nebular (reflection) origin for this sky spectrum
component.

In Tr~14/16, the absence of scattered solar continuum enables us to
eliminate all time-variable sky signatures by just subtracting out the
sky-glow lines: this will leave a purely "star plus nebular" spectrum,
and the nebular part will be later estimated from the whole set of
(non-simultaneous) sky spectra in the same dataset.  Sky-glow lines
in the HR15n range, of sufficient intensity as to merit consideration
in this context, are no more than 20 narrow lines. In principle, also
some geocoronal \ha\ emission is expected, but in practice it is not
detected against the enormously brighter \ha\ emission from the nebula.
For each Observing Block, the sky-glow line spectrum, net of the adjacent
continuum intensity, and averaged within the same Observing Block, was
computed, and subtracted out from all spectra in the same Observing
Block to obtain "star plus nebular" or "nebular-only" spectra, for
stellar and sky spectra respectively.

Having eliminated terrestrial sky features in this way, the problem now
reduces to estimate the most appropriate nebular spectrum to be used
for correcting a given stellar spectrum.  Since the spatial density of
sky fibres (considering all Observing Blocks together) is much less than
that of target star fibres, the correspondence between the nebular (and
sky-continuum) emission in a stellar spectrum and in its nearest-neighbor
sky spectrum is never found to be perfect.  Distances between a given
star and its nearest sky fibre are of order of 30" or more: the nebular
morphology is here so complex that there is no guarantee that even the
nearest sky is a good approximation to the sought nebular spectrum
at the star position; one gets an idea of the range of variation of
nebular spectrum in the vicinity of a given star by looking at more
(say five) sky positions nearby. Therefore, we made subtraction of
nebular spectra from stellar spectra five times per star, using its
nearest five sky positions. Since the properties of nebular emission
vary sometimes sharply, and definitely non-linearly in space, this is
considered a robust method (within the limitations of the available data)
to understand the uncertainties involved in the procedure.

}

\end{appendix}

\end{document}